\documentclass[aps,prx,preprint,groupedaddress]{revtex4-2}
\usepackage{graphicx}
\usepackage{bm}

\begin{document}


\title{Berry connection from many-body wave functions and superconductivity: Calculations by the particle number conserving Bogoliubov-de Gennes equations}


\author{Hiroyasu Koizumi}
\affiliation{Center for computational sciences, University of Tsukuba, Tsukuba, Ibaraki, Japan}
\email{koizumi.hiroyasu.fn@u.tsukuba.ac.jp}
\author{Alto Ishikawa}
\affiliation
{Graduate School of Pure and Applied Sciences, University of Tsukuba, Tsukuba, Ibaraki, Japan}


\date{\today}

\begin{abstract}
A fundamentally revised version of superconductivity theory has been put forward by the present authors since the standard theory of superconductivity based on the BCS theory cannot explain superconductivity in cuprates discovered in 1986, and reexaminations on several experimental results on the conventional superconductors indicate the necessity for a fundamental revision.

The revision is made on the origin of the superconducting phase variable, which is attributed to a Berry connection arising from many-body wave functions. With this revision, the theory can be cast into a particle number conserving formalism.

We have developed a method to calculate superconducting states with the Berry connection 
using the particle number conserving version of the Bogoliubov-de Gennes equations.
An example calculation is made for a model originally built for cuprate superconductors.
\end{abstract}


\maketitle

\section{Introduction}

Since the present work deals with a fundamental revision of the standard theory of superconductivity, we start with
historical evolution of superconductivity theory.

 After studying the gauge theories in quantum mechanics \cite{London1927}, 
 London developed a phenomenological theory of superconductivity from the view point of the wave mechanics. He put forward the equation 
\begin{eqnarray}
\hbar \nabla \chi^{\rm super}=m {\bf v}+{ q \over c}{\bf A}^{\rm em}
\label{Londoneq4b}
\end{eqnarray}
which explains the Meissner effect \cite{London1950}, where $\hbar$ is Planck's constant divided by $2\pi$, $\chi^{\rm super}$ is the superpotential, $q$ and $m$ are the charge and mass of the particle, $c$ is the speed of light in vacuum, and ${\bf A}^{\rm em}$ is the electromagnetic vector potential, respectively.

London argued that the wave function for the superconducting state is rigid in the sense that
when a magnetic field is applied, and the wave function is modified only by the change of the phase factor
\begin{eqnarray}
\Psi_0 \rightarrow \Psi=\Psi_0 e^{{ i } \sum_j \chi^{\rm super} ({\bf r}_j)}
\label{LondonWF}
\end{eqnarray}
where $\Psi_0$ is the wave function when magnetic field is absent, and  $\Psi$ is that when it is present \cite{London1950}.

The supercurrent density ${\bf j}_s$ can be obtained using the rigid wave function in Eq.~(\ref{LondonWF}) as
\begin{eqnarray}
{\bf j}_s=qn_s{\bf v}=-{{q^2 n_s} \over {mc}} \left({\bf A}^{\rm em} -{{\hbar c} \over q} \nabla \chi^{\rm super}
\right)
\label{Londoneq1}
\end{eqnarray}
where $n_s$ is the number density of the charge carriers. This explains the diamagnetic current that flows when a magnetic field is applied.

When Eq.~(\ref{Londoneq1}) is combined with one of Maxwell's equation 
\begin{eqnarray}
\nabla \times {\bf B}^{\rm em}={{4\pi} \over c}{\bf j}_s
\end{eqnarray}
 the equation that
explains the exclusion of magnetic field from inside superconductors (Meissner effect),
\begin{eqnarray}
\nabla^2 {\bf B}^{\rm em}=-{ 1 \over {\lambda_L^2}} {\bf B}^{\rm em}
\end{eqnarray}
 is obtained, where the London penetration depth
\begin{eqnarray}
\lambda_L=\sqrt{{mc^2} \over {4 \pi q^2 n_s}}
\label{lm-L}
\end{eqnarray}
was introduced. This is the length scale where a magnetic field penetrates into a superconductor.

 The superpotential gives rise to fluxoid,
\begin{eqnarray}
\Phi_{\rm fluxoid}={{c \hbar} \over q } \oint \nabla \chi^{\rm super} \cdot d{\bf r}
\label{eqFluxoid0}
\end{eqnarray}

In a ring-shaped superconductor where the magnetic field is excluded from the bulk, 
the flux through the ring is given by
\begin{eqnarray}
\oint_C {\bf A}^{\rm em} \cdot d{\bf r} =\Phi_{\rm fluxoid}
\label{eqFluxoid}
\end{eqnarray}
where $C$ is a loop along the ring inside the superconductor where ${\bf j}_s=0$.

Experiments confirmed the presence of the fluxoid given by
\begin{eqnarray}
\Phi_{\rm fluxoid}={ {hc} \over {2e}}n
\label{eqDeaver}
\end{eqnarray}
where $n$ is an integer \cite{Deaver1961}.

Note that
\begin{eqnarray}
   {\bf A}^{\rm eff}= {\bf A}^{\rm em}-{{\hbar c} \over q}\nabla \chi^{\rm super} 
   \label{eqAeff}
 \end{eqnarray}
 must be gauge invariant since ${\bf j}_s$ in Eq.~(\ref{Londoneq1}) is gauge invariant. Thus, the angular variable $\chi^{\rm super}$ that
 makes ${\bf A}^{\rm eff}$ gauge invariant exists in superconductors.

The theory based on the wave mechanics formalism did not succeed in explaining superconductivity. The first successful microscopic theory for superconductivity, the BCS theory, was built by the formalism using creation and annihilation operators \cite{BCS1957}.
  
Let $c^{\dagger}_{{\bf k} \sigma}$ be the creation operator for the electron in the metal with wave vector ${\bf k}$ and spin $\sigma$, 
the BCS theory uses the ground state composed of different particle number states
\begin{eqnarray}
|{\rm BCS} (\theta) \rangle=\prod_{\bf k}\left(u_{\bf k}+v_{\bf k}c^{\dagger}_{{\bf k} \uparrow}c^{\dagger}_{-{\bf k} \downarrow}
e^{ {i}{\theta}} \right)|{\rm vac} \rangle
\label{BCS}
\end{eqnarray}
where $u_{\bf k}$ and $v_{\bf k}$ are real parameters that satisfy $u_{\bf k}^2+v_{\bf k}^2=1$, and $|{\rm vac} \rangle$ is the vacuum that satisfy
\begin{eqnarray}
c_{{\bf k} \sigma}|{\rm vac} \rangle=0
\label{vacuum}
\end{eqnarray}

The ground state $|{\rm BCS} (\theta) \rangle$ is a linear combination of different particle number states, thus, breaks the conservation of the particle number. It has a degeneracy with respect to the choice of $\theta$, and breaking the global $U(1)$ gauge invariance,  
the invariance of the physical state by the change $c_{{\bf k} \sigma} \rightarrow e^{-{i \over 2}\theta}c_{{\bf k} \sigma}, c^{\dagger}_{{\bf k} \sigma} \rightarrow e^{{i \over 2}\theta}c^{\dagger}_{{\bf k} \sigma}$, where $\theta$ is a constant.

Using the particle number non-conserving formalism, the BCS theory provides a way to calculate the superconducting transition temperature as the energy gap formation temperature by the electron-pairing.

A salient feature of the BCS theory is the presence of the Bogoliubov excitation with an energy gap formed by the electrton-pairing \cite{Bogoliubov58}.
This excitation is most clearly seen if we use the Bogoliubov operators given by
\begin{eqnarray}
 \gamma_{{\bf k} \uparrow }&=& u_{k} e^{-{i \over 2}\theta} c_{{\bf k} \uparrow} -  v_{ k} e^{{i \over 2}\theta} c^{\dagger}_{-{\bf k} \downarrow}
\nonumber
\\
\gamma_{-{\bf k} \downarrow }&=& u_{k} e^{-{i \over 2}\theta} c_{-{\bf k} \downarrow} +  v_{ k} e^{{i \over 2}\theta} c^{\dagger}_{{\bf k} \uparrow}
\label{eqBogoliubov58}
\end{eqnarray}
where they satisfy
\begin{eqnarray}
 \gamma_{{\bf k} \uparrow }|{\rm BCS} (\theta) \rangle=0, \quad \gamma_{-{\bf k} \downarrow }|{\rm BCS} (\theta) \rangle=0
 \end{eqnarray}
 This indicates that the superconducting state is the ``vacuum of the Bogoliubov quasiparticlres'', which replaces Eq.~(\ref{vacuum}).
 
 Using the Bogoliubov operators, the BCS Hamiltonian can be cast into the following, 
\begin{eqnarray}
H_{\rm BCS}=\sum_{{\bf k}, \sigma}  E_{\bf k}\gamma^{\dagger}_{{\bf k} \sigma } \gamma_{{\bf k} \sigma }+E_{\rm const}
 \end{eqnarray}
 where 
 \begin{eqnarray}
 E_{\bf k}=\sqrt{ ({\cal E}({\bf k}))^2+\Delta_{\bf k}^2}
  \end{eqnarray}
  with ${\cal E}({\bf k})$ being the energy of Bloch electrons measured from the Fermi energy;
   $\Delta_{\bf k}$ is the energy gap given by
  \begin{eqnarray}
 \Delta_{\bf k}=2E_{\bf k} u_k v_k, \quad u_k^2={1 \over 2} \left(1 +  {{{\cal E}({\bf k})} \over {E_{\bf k}}}   \right), \quad v_k^2={1 \over 2} \left(1    - {{{\cal E}({\bf k})} \over {E_{\bf k}}}   \right)
 \end{eqnarray}
$E_{\bf k}$ explains the energy gap spectrum observed in superconductors.

In the BCS theory, the induced current by a magnetic field is calculated as a linear response to ${\bf A}^{\rm em}$, which gives the formula
very similar to Eq.~(\ref{Londoneq1}) without the term with $\nabla \chi^{\rm super}$. The obtained current is not gauge invariant, and the
gauge invariance of the induced current became a big issue. After seminal works by Anderson \cite{Anderson1958a,Anderson1958b}, a final resolution was provided by Nambu\cite{Nambu1960}, culminating to the idea ``spontaneous gauge symmetry breaking''.
Nambu showed the the Nambu-Goldstone mode appears due to the degeneracy of the ground state with respect to the choice of $\theta$, 
and this mode retrieves the gauge invariance. 

Independent of the BCS theory, a phenomenological theory was developed by Ginzburg and Landau \cite{GL} based in the London theory of superconductivity.
The gauge invariant supercurrent is most easily calculated by this Ginzburg-Landau theory \cite{GL}. 
In this theory a free energy for the superconducting state is given as a functional of an order parameter 
 \begin{eqnarray}
 \Psi_{\rm GL}=n_s^{1/2}e^{ i \chi^{\rm super}}
 \end{eqnarray}
 
The free energy is composed of a material part
 \begin{eqnarray}
 F_{\rm mat}=F_{\rm normal}+
 \int d^3 r {1 \over {2m}}  \left| \left( { \hbar \over i} \nabla -{q \over c}{\bf A}^{\rm em} \right) \Psi_{\rm GL} \right|^2
 +  \int d^3 r  \left( \alpha  |\Psi_{\rm GL}|^2 +{ \beta \over 2}|\Psi_{\rm GL}|^4 \right)
 \nonumber
 \\
 \label{mat1}
\end{eqnarray}
where $\alpha <0 $, $\beta >0$ are parameters and $F_{\rm normal}$ is the free energy of the normal phase,
and a magnetic field part 
\begin{eqnarray}
 F_{\rm mag}=\int d^3 r {1 \over {8\pi}} \left( {\bf B}^{\rm em}  \right)^2 
  \label{mag}
 \end{eqnarray}
  
In the Ginzburg-Landau theory, the supercurrent is calculated as
\begin{eqnarray}
{\bf j}_s=-c{{\delta F_{\rm mat}} \over {\delta {\bf A}^{\rm em}({\bf r})}}=-{{q^2 n_s} \over {mc}} \left({\bf A}^{\rm em} -{{\hbar c} \over q} \nabla \chi^{\rm super} \right)
\end{eqnarray}
in accordance with Eq.~(\ref{Londoneq1}). The gauge invariance is retrieved by $\chi^{\rm super}$ \cite{Weinberg}. 

The connection between the BCS and Ginzburg-Landau theories was provided by Gor'kov \cite{Gorkov1959}.
The Gor'kov's derivation yields,
\begin{eqnarray}
m=2m^{\ast}, \quad q=-2e
\end{eqnarray}
where $m^{\ast}$ is the effective mass of electron in the metal. 

Although the  Gor'kov derivation has been considered as a valid one,
it disagrees with the experimental value $m=2m_e$ from the London moment measreuemen where $m_e$ is the free electron mass \cite{Hirsch2013b,koizumi2021}.

A notable point of the Ginzburg-Landau theory is that it introduces the coherence length
\begin{eqnarray}
\xi_{\rm GL}=\sqrt{ \hbar^2 \over {2m|\alpha|}}
 \label{xi-GL}
\end{eqnarray}

The BCS theory also introduces a similar quantity,  the BCS coherence length,
 \begin{eqnarray}
 \xi_{\rm BCS}= {{ \hbar v_{\rm Fermi}} \over {\pi \Delta}}
 \label{xi-BCS}
 \end{eqnarray}
 Currently, it is interpreted that it is the size of the electron-pair, where $\Delta$ is the energy gap and $ v_{\rm Fermi}$ is the velocity of the electron at the Fermi energy \cite{BCS1957}.  
 
Based on the Ginzburg-Landau theory, Abrikosov showed that there are two types of superconductors depending on the ratio of $\lambda_L$ and $\xi_{\rm GL}$,
\begin{eqnarray}
\mbox{type I     } \quad {\lambda_L \over {\xi_{\rm GL}}} < { 1 \over \sqrt{2}}; \quad \quad \mbox{type II    } \quad {\lambda_L \over {\xi_{\rm GL}}} > { 1 \over \sqrt{2}}, 
\end{eqnarray}
In type II superconductors, applied magnetic field can penetrate the sample as vortices, and $\xi_{\rm GL}$ is the core size of such vortices \cite{Abrikosov}.

\begin{figure}
\begin{center}
\includegraphics[scale=0.5]{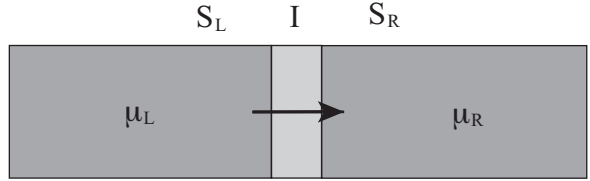}
\end{center}
\caption{Josephson junction S$_{\rm L}$-I-S$_{\rm R}$ (superconductor-insulator-superconductor junction) considered by Josephson. $\mu_R$ and $\mu_L$ are chemical potentials of S$_{\rm L}$ and S$_{\rm R}$, respectively. When voltage $V$ exists the relation $eV=\mu_R-\mu_L$ holds.}
\label{Josephson2}
\end{figure}

Remarkable effects caused by $\chi^{\rm super}$ were predicted by Josephson \cite{Josephson62}. They are called ``dc Josephson effect'' and
``ac Josephson effect'', and 
 occur in the junction composed of two superconductors separated by a thin barrier (or insulator region) depicted in Fig.~\ref{Josephson2}.
 
 The current through the junction is given by
 \begin{eqnarray}
J=J_c \sin \phi
\end{eqnarray}
where $J_c$ is a parameter for the junction, and $\phi$ is given using the gauge invariant ${\bf A}^{\rm eff}$ as
\begin{eqnarray}
\phi=-{ q \over {\hbar c}} \int_L^R \left( {\bf A}^{\rm em}-{{\hbar c} \over q}\nabla \chi^{\rm super} \right)+\phi_0
\label{eqphi}
\end{eqnarray}
where $\phi_0$ is a constant and the integration is taken along the line connecting the two superconductors.

The ac Josephson effect occurs when there is a voltage across the junction. 
When the voltage is $V$, time variation of $\phi$ given by
\begin{eqnarray}
\dot{\phi}={{2eV} \over \hbar}
\label{eq2e}
\end{eqnarray}
occurs.

This effect is most clearly seen if  a radiation field with frequency $f$ exists;
the resonance condition $\dot{\phi}=2\pi f n$, where $n$ is an integer, gives
\begin{eqnarray}
V={{h f} \over {2e}}n
\label{eqShapiro}
\end{eqnarray}
Then, steps are observed in the $I$-$V$ curve (current-voltage curve) \cite{Shapiro63}.

 Josephson obtained Eq.~(\ref{eq2e}) based on the BCS theory by assuming $q=-2e$ \cite{Josephson62}.
His derivation had been long considered to be right one. However, it has been argued that it misses a contribution; Eq.~(\ref{eq2e}) can be obtained by $q=-e$ \cite{Koizumi2011,koizumi2020c,koizumi2020,HKoizumi2015,koizumi2021b}.

Before the discovery of high temperature superconductivity in 1986 \cite{Muller1986}, 
the BCS theory and its extensions have been very successful and was believed that superconductivity was a well-understood solved problem. 
However, the cuprate superconductivity is markedly different from the BCS superconductivity in a number of ways;
 for example, 
the magnetism is harmful in the BCS superconductivity since it causes the breaking of electron-pairing, however magnetism coexists with superconductivity in the cuprate \cite{Kastner1998};
the superconducting transition temperature is not the energy gap formation temperature, but corresponds to the stabilizing temperature for loop currents of $\xi_{\rm GL}$ size in optimally doped sample of the cuprates \cite{Kivelson95};
the charge carriers in the room temperature become small polarons at low temperatures \cite{Bianconi}.

Since the cuprate superconductivity is so different from the BCS superconductivity, 
a theory much departs from the standard one is expected to be needed \cite{Koizumi2011}. 
The superconducting state transition temperature indicates the importance of $\xi_{\rm GL}$-sized loop currents, and
it has been argued that $\xi_{\rm GL}$-sized spin-vortices and loop currents appear around the small polarons in the bulk, explaining the
magnetic excitation spectrum observed by inelastic neutron scattering \cite{Hidekata2011}.

Efforts toward the elucidation of the cuprate superconductivity lead some researchers to reexamine superconductivity from very fundamental levels. Those efforts revealed that there are some experimental facts disagreeing with the standard theory even in the conventional superconductors \cite{Hirsch2009}. 
We would like to point out four major problems below:
 \vspace{5 mm}

 1) Supercurrent in the standard theory contradicts the reversible superconducting-normal phase transition in a magnetic field observed in type I superconductors.
 
 \vspace{2 mm}
 
Magnetic field is excluded from the bulk of superconductors, thereby the supercurrent is generated in the surface region of width $\lambda_L$ in Eq.~(\ref{lm-L}) as the screening current for the magnetic field.  

However, the Meissner effect is not just the exclusion of the magnetic field. It also indicates that the superconducting state is a thermodynamically stable phase in the $T-H$ plane, where $T$ is the temperature and $H$ is the external magnetic field.  
Thus, the surface current stops without generating irreversible Joule heat during
the superconducting to currentless normal metallic phase transition. Note that the thermodynamically stable phase of the normal metal in a magnetic field is currentless.

Actually, after the discovery of the Meissner effect, the reversibility of the superconducting-normal state transition in the presence of a magnetic field was confirmed \cite{Keesom1,Keesom2,Keesom3, Keesom4,Keesom}; and the state of the art calorimetry indicates that 99.99\% of the supercurrent stops without current carriers undergoes irreversible collisions (see Appendix B of Ref.~\cite{Hirsch2017}). 
Hirsch argued that the standard theory of superconductivity cannot explain how this transition is possible since the supercurrent generated by the flow of electron pairs inevitably produces the Joule heat due to the existence of a significant number of  broken pairs that flow with dissipation during the transition \cite{Hirsch2017,Hirsch2018,Hirsch2020}. 

 It is also note worthy that the use of the linear response theory to calculate the supercurrent may be problematic since it is tied to the fluctuation-dissipation relation \cite{Kubo1957}, and the current calculated by it usually causes dissipation. This point has drawn renewed attention in relation to the
 absence of a dissipative quantum phase transition in Josephson junctions \cite{PhysRevX2021a,PhysRevX2021b,PhysRevX2021c}.
 
 One may say that the
 Ginzburg-Landau theory is derivable from the BCS theory \cite{Gorkov1959}, the supercurrent generation in the latter theory 
 can be used as the evidence that the BCS theory explains the Meissner effect.
 However, the Gor'kov's derivation from BCS to Ginzburg-Landau theories disagrees in the mass in the London moment as explained below.
  
  \vspace{5 mm}
 
 2) Experiments indicate that the mass in the London moment is the free electron mass, however, the standard theory predicts it to be an effective mass.
 
 \vspace{2 mm}
 
 The London moment is the magnetic moment generated by a rotating superconductor in a magnetic field  \cite{Becker1933,London1950}. Inside the superconductor magnetic field, the ``{\em London field}'', is produced by the supercurrent flowing in the surface region. 
 
 Let us consider a superconducting sphere rotating about its symmetry axis with constant angular velocity ${\bm \omega}$.
Then, the velocity at the position ${\bf r}$ measured from  the center of the sphere is given by
\begin{eqnarray}
{\bf v}={\bm \omega} \times {\bf r}
\end{eqnarray}

Electrons inside the superconductor move with this velocity to shield the background positive charge.
Substituting this in Eq.~(\ref{Londoneq4b})  and using the relation ${\bf B}^{\rm em}=\nabla \times {\bf A}^{\rm em}$, we have
\begin{eqnarray}
{\bf B}^{\rm em}=-{{2m c} \over q} {\bm \omega}
\end{eqnarray}
This is the London field, where $m$ is mass, $q$ is charge, and $c$ is the speed of light.

The London moment and field have been measured many times using different materials, ranging from the conventional superconductor \cite{Hildebrandt1964,Zimmerman1965,Brickman1969,Tate1989,Tate1990} to the high T$_{\rm c}$ cuprates \cite{VERHEIJEN1990a,Verheijen1990} and
heavy fermion superconductors \cite{Sanzari1996}. The results always indicate that the mass $m$ 
is the free electron mass $m_e$ if $q=-e$ is used, not the effective mass $m^{\ast}$ predicted by the standard theory including the Gorkov's derivation.

 \vspace{5 mm}

 3) In the standard theory, the breakdown of the global $U(1)$ gauge invariance or the non-conservation of the particle number is essential.
 
 \vspace{2 mm}
  
 It is sensible to consider that the particle number is conserved in an isolated superconductor. However, the standard theory requires to use 
 the particle number non-conserving formalism as manifested in the use of Eq.~(\ref{BCS}) \cite{Peierls1991,LeggettBook}.
 It has been also claimed that the use of $\theta$ as a physically meaningful parameter is against the superselection rule for charge \cite{WWW1970}. 
 Thus, it has been argued that current formalism is an approximation of the true formalism in which the particle number is conserved \cite{Peierls92,LeggettBook}.
 
 \vspace{5 mm}
 
 4) The derivation of the ac Josephson effect takes into account only the half of the contributions.
  
 \vspace{2 mm}
 
 We shall revisit the ac Josephson effect: let us calculate $\dot{\phi}$ using Eq.~(\ref{eqphi}),
\begin{eqnarray}
 \dot{\phi}=-{q \over {\hbar c}} \int_R^L d{\bf r} \cdot \left(\partial_t{\bf A}^{\rm em} -{{\hbar c} \over {q}} \nabla \partial_t{\chi}^{\rm super} \right)
 ={q \over \hbar} \int_R^L d{\bf r} \cdot {\bf E}^{\rm em}  -\left. {q \over \hbar}  \left( - \varphi^{\rm em} -{\hbar \over {q}}  \partial_t{\chi}^{\rm super} \right)\right|^L_R
 \nonumber
 \\
 \label{eqnPhidot}
 \end{eqnarray}
 where the relation
\begin{eqnarray}
 {\bf E}^{\rm em} =-{1 \over c}\partial_t{\bf A}^{\rm em} - \nabla \varphi^{\rm em}
 \end{eqnarray}
 is used.

  \begin{figure}
\begin{center}
\includegraphics[scale=0.6]{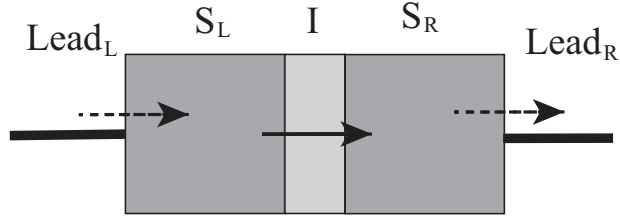}
\end{center}
\caption{Josephson junction in a real experimental situation, connected to leads, denoted as Lead$_{\rm L}$ and Lead$_{\rm R}$. 
The ac Josephson effect is observed as the quantized voltage given in Eq.~(\ref{eqShapiro}) under the application a microwave of frequency $f$ and a dc current through the junction.}
\label{Josephson}
\end{figure}
 
 There are two contributions for $\dot{\phi}$.
 The first one is
 \begin{eqnarray}
 {q\over \hbar} \int_R^L d{\bf r} \cdot {\bf E}^{\rm em} =-{{q V} \over \hbar}
  \end{eqnarray}
 where $V$ 
   is the voltage across the junction.
   
   The second one is
 \begin{eqnarray}
    -\left. {q \over \hbar}  \left( - \varphi^{\rm em} -{\hbar \over {q}}  \partial_t{\chi}^{\rm super} \right)\right|^L_R
    ={q \over \hbar}  {{\mu_L-\mu_R } \over e}
\end{eqnarray}
Here, the time-component partner of Eq.~(\ref{eqAeff}), 
     \begin{eqnarray}
  \varphi^{\rm eff}= \varphi^{\rm em} +{\hbar \over {q}}  \partial^{\rm super}_t{\chi} 
  \end{eqnarray}
  is identified as 
 \begin{eqnarray}
 \varphi^{\rm eff}={ \mu  \over e }
   \end{eqnarray}
  where $\mu$ is the chemical potential.
   
   The balance between the voltage and the chemical potential difference yields,
  \begin{eqnarray}
 eV =\mu_R-\mu_L 
 \label{eqBalance}
 \end{eqnarray}
 
  Thus, the sum of the two contributions yields
\begin{eqnarray}
\dot{\phi}=-{{2qV} \over \hbar}
\label{eqJR}
\end{eqnarray}
In order to obtain the result in Eq.~(\ref{eq2e}), $q$ needs to be $q=-e$, this contradicts the value used by Josephson $q=-2e$.

The above discrepancy comes from the fact that only one of the two contributions in Eq.~(\ref{eqnPhidot}) is included in the Josephson's derivation \cite{Josephson62,Josephson1965}.

The first term with ${\bf E}^{\rm em}$ in Eq.~(\ref{eqnPhidot}) arises when charged particles go through the electric field in the insulator region.
The second term with $\varphi^{\rm em}$ and $\chi^{\rm super}$ arises due to the fact that the chemical potentials are maintained by the contact with the leads; in order to keep the chemical potential constant, electrons enter from one of the leads and exit to the other lead as electrons tunnel through ``I'' region in Fig.~\ref{Josephson}.
Since experiments are performed in the presence of a dc current in addition to a microwave field, it is sensible to
consider that the two contributions both exit.

If $q=-2e$ is used as in the standard theory, the Josephson relation becomes $\dot{\phi}={{4eV} \over \hbar}$. In this case, half-integers will appear in Eq.~(\ref{eqShapiro}). Indeed, such effects have been observed \cite{Ueda2020}. 

As Eq.~(\ref{Londoneq1}) indicates,
the most important ingredient for supercurrent generation is $\chi^{\rm super}$. If an angular variable$\chi$ with period $2\pi$ that satisfy
\begin{eqnarray}
\nabla \chi^{\rm super}=-{1 \over 2} \nabla \chi
\label{Londoneq10}
\end{eqnarray}
exits, the experimental values of fluxoid are obtained by $q=-e$. 
$q=-2e$ contributions may exist since the
electron pairing occurs in the system; however, the view that the supercurrent is the flow of paired-electrons will not be valid according to the
problem given above in the item 1). It is also noteworthy to recognize that careful treatments of the spatial and temporal fluctuations of ${\bf E}^{\rm em}$ are needed to understand Josephson effect related phenomena, correctly \cite{zhang2021,koizumi2021b}.
 
The new theory is the one that can be formulated in a particle number conserving manner \cite{koizumi2019,koizumi2020,koizumi2020b,koizumi2020c,koizumi2021,koizumi2021b}. We will  present a method to calculate properties of superconducting states, such as diamagnetic response current and persistent current, using the particle number conserving formalism.

The organization of the present work is as follows:
 In Section~\ref{sec4} particle number conserving version of the BCS theory is explained. This formalism becomes possible due to
 the presence of non-trivial Berry connection. 
 In Section~\ref{sec5} model calculations are performed using the particle number conserving Bogoliubov-de Gennes equations derived from 
 the particle number conserving version of the BCS theory. A model constructed for the cuprate superconductivity is used to demonstrate calculations. 
  We conclude this work with some remarks in Section~\ref{sec6}.
  
  Four appendices are included: Berry connection from many-body wave functions and the modification of Maxwell's equations are explained in
  Appendix~\ref{Appendix1};
 appearance of ${\bf A}^{\rm fic}$ from spin-twisting itinerant motion of electrons is explained in Appendix~\ref{Appendix2}; 
 reversible superconducting-normal metal phase transition in a magnetic field is discussed in Appendix~\ref{Appendix3}; and 
  collective mode $\chi$ and associated number changing operators are derived in Appendix~\ref{Appendix4}.

 \section{Particle number conserving version of the BCS theory using the superconducting phase from the Berry connection}
 \label{sec4}
 
 In the original BCS theory, electrons in the normal metallic state are assumed to be well-described by the free electrons with the effective mass $m^{\ast}$.
 Then, the electron field operators are given by
\begin{eqnarray}
\hat{\Psi}_{\sigma}({\bf r})={ 1 \over \sqrt{\cal V}}\sum_{\bf k} e^{i {\bf k} \cdot {\bf r}} c_{{\bf k} \sigma}
\end{eqnarray}
where ${\cal V}$ is the volume of the system.

Using the Bogoliubov operators \cite{deGennes,Zhu2016} they are expressed as
\begin{eqnarray}
\hat{\Psi}_{\uparrow}({\bf r})&=&{ 1 \over \sqrt{\cal V}}\sum_{\bf k} e^{{i \over 2}\theta}e^{i {\bf k} \cdot {\bf r}} \left( \gamma_{{\bf k} \uparrow } u_{ k} -  \gamma^{\dagger}_{-{\bf k} \downarrow } v_{k} \right)
\nonumber
\\
\hat{\Psi}_{\downarrow}({\bf r})&=&{ 1 \over \sqrt{\cal V}}\sum_{\bf k} e^{{i \over 2}\theta}e^{i {\bf k} \cdot {\bf r}} \left( \gamma_{{\bf k} \downarrow } u_{ k} +\gamma^{\dagger}_{-{\bf k} \uparrow } v_{ k} \right)
\label{FieldOp}
\end{eqnarray}

Now we depart from the standard theory by replacing $e^{\pm{i \over 2}\theta}$ in Eq.~(\ref{FieldOp}) by the number changing operators $e^{\pm {i \over 2}\hat{\chi}}$. Here, we do not specify the coordinate in $\hat{\chi}$ (we will introduce the coordinate dependence, later). 
Then, the Bogoliubov transformation in Eq.~(\ref{eqBogoliubov58}) becomes
\begin{eqnarray}
 \gamma_{{\bf k} \uparrow }&=& u_{k} e^{{i \over 2}\hat{\chi}} c_{{\bf k} \uparrow} -  v_{ k} e^{-{i \over 2}\hat{\chi}} c^{\dagger}_{-{\bf k} \downarrow}
\nonumber
\\
\gamma_{-{\bf k} \downarrow }&=& u_{k} e^{{i \over 2}\hat{\chi}} c_{-{\bf k} \downarrow} +  v_{ k} e^{-{i \over 2}\hat{\chi}} c^{\dagger}_{{\bf k} \uparrow}
\end{eqnarray}

Note that the above Bogoliubov operators conserve particle numbers.
Terms like $e^{{i \over 2}\hat{\chi}} c_{{\bf k} \sigma}$ can be interpreted that an electron in the $({\bf k}, \sigma)$ single-electron mode is 
removed and an electron is added to the collective mode described by $\chi$; those like $e^{-{i \over 2}\hat{\chi}} c^{\dagger}_{{\bf k} \sigma}$ create an electron in the $({\bf k}, \sigma)$ single-electron mode and subtract an electron from the collective mode described by $\chi$. Thus, the Bogoliubov operators cause the fluctuation of the number of electrons in the collective mode by transferring electrons between single-particle modes and collective mode. As a consequence, the ground state becomes a linear combination of states with different number of electrons in the collective mode \cite{koizumi2019}.
This state replaces the state with fluctuating total number of electrons (given by Eq.~(\ref{BCS})) in the standard theory.

By including ${1 \over \sqrt{\cal V}}e^{ i {\bf k} \cdot {\bf r}}$ in $u_k$ and $v_k$, we define
\begin{eqnarray}
u_{\bf k}({\bf r})={1 \over \sqrt{\cal V}}e^{ i {\bf k} \cdot {\bf r}}u_k, \quad
v_{\bf k}({\bf r})={1 \over \sqrt{\cal V}}e^{i {\bf k} \cdot {\bf r}}v_k
\end{eqnarray}

Then, the field operators become
\begin{eqnarray}
\hat{\Psi}_{\uparrow}({\bf r})&=&\sum_{\bf k} e^{-{i \over 2}\hat{\chi}({\bf r})} \left( \gamma_{{\bf k} \uparrow } u_{\bf k}({\bf r})  -\gamma^{\dagger}_{{\bf k} \downarrow } v^{\ast}_{\bf k}({\bf r}) \right)
\nonumber
\\
\hat{\Psi}_{\downarrow}({\bf r})&=&
\sum_{\bf k} e^{-{i \over 2}\hat{\chi}({\bf r})}\left( \gamma_{{\bf k} \downarrow } u_{\bf k}({\bf r}) +\gamma^{\dagger}_{{\bf k} \uparrow } v^{\ast}_{\bf k}({\bf r}) \right)
\end{eqnarray}
where the spatial dependence is included in $\hat{\chi}$.

 Now we allow the coordinate dependent functions other than plane waves. 
 We use the label $n$ in place of the wave number ${\bf k}$ for them. 
 As a result, the field operators become
\begin{eqnarray}
\hat{\Psi}_{\uparrow}({\bf r})&=&\sum_{n} e^{-{i \over 2}\hat{\chi} ({\bf r})}\left( \gamma_{{n} \uparrow } u_{n}({\bf r})  -\gamma^{\dagger}_{{n} \downarrow } v^{\ast}_{n}({\bf r}) \right)
\nonumber
\\
\hat{\Psi}_{\downarrow}({\bf r})&=&
\sum_{n} e^{-{i \over 2}\hat{\chi} ({\bf r})} \left( \gamma_{{n} \downarrow } u_{n}({\bf r}) +\gamma^{\dagger}_{{n} \uparrow } v^{\ast}_{n}({\bf r}) \right)
\end{eqnarray}

The particle number conserving Bogoliubov operator $\gamma_{n \sigma}$ satisfies
\begin{eqnarray}
\gamma_{n \sigma}|{\rm Gnd}(N) \rangle=0
\end{eqnarray}
where $|{\rm Gnd}(N) \rangle$ is the ground state with the total number of particles $N$.

The ground state also satisfies
\begin{eqnarray}
e^{\pm {i \over 2} \hat{\chi}({\bf r})}|{\rm Gnd}(N) \rangle= e^{\pm{i \over 2} {\chi}({\bf r})}|{\rm Gnd}(N\pm1) \rangle
\label{eqBC}
\end{eqnarray}

The above phase factor gives
\begin{eqnarray}
\langle {\rm Gnd}(N)|e^{ {i \over 2} \hat{\chi}({\bf r}_f)}e^{-{i \over 2} \hat{\chi}({\bf r}_i)}|{\rm Gnd}(N) \rangle
=e^{ {i \over 2} \int_{{\bf r}_i}^{{\bf r}_f} \nabla {\chi} \cdot d{\bf r}}
\end{eqnarray}
indicating that the phase ${1 \over 2} \int_{{\bf r}_i}^{{\bf r}_f} \nabla {\chi} \cdot d{\bf r}$ is acquired when the particle travels from ${\bf r}_i$ to ${\bf r}_f$,
in accordance with the presence of ${\bf A}^{\rm fic}$ in Eq.~(\ref{eqS2'}).

Let us consider the electronic Hamiltonian expressed by the field operators
\begin{eqnarray}
H=\sum_{\sigma} \int d^3 r \hat{\Psi}^{\dagger}_{\sigma}({\bf r}) h({\bf r}) \hat{\Psi}_{\sigma}({\bf r}) 
-{1 \over 2} \sum_{\sigma, \sigma'}\int d^3 r d^3 r' V_{\rm eff}({\bf r}, {\bf r}') \hat{\Psi}^{\dagger}_{\sigma}({\bf r}) \hat{\Psi}^{\dagger}_{\sigma'}({\bf r}') \hat{\Psi}_{\sigma'}({\bf r}') \hat{\Psi}_{\sigma}({\bf r}) 
\end{eqnarray}
where $h({\bf r})$ is the single-particle Hamiltonian given by
\begin{eqnarray}
h({\bf r})={ 1 \over {2m_e}} \left( { \hbar \over i} \nabla +{e \over c} {\bf A}^{\rm em} \right)^2+U({\bf r})-\mu 
\end{eqnarray}
and $U({\bf r})$ is the single particle potential energy, $\mu$ is the chemical potential, and $-V_{\rm eff}$ is the effective interaction between electrons. In the BCS theory, $-V_{\rm eff}$ describes the attractive interaction between electrons
via the virtual phonon exchange.

We perform the following mean field approximation 
\begin{eqnarray}
H^{\rm MF}&=&\sum_{\sigma} \int d^3 r \hat{\Psi}^{\dagger}_{\sigma}({\bf r}) h({\bf r}) \hat{\Psi}_{\sigma}({\bf r}) 
+\int d^3 r d^3 r' 
\left[ \Delta({\bf r}, {\bf r}')\hat{\Psi}^{\dagger}_{\uparrow}({\bf r}) \hat{\Psi}^{\dagger}_{\downarrow}({\bf r}') e^{-{i \over 2}(\hat{\chi}({\bf r}) +\hat{\chi}({\bf r}')) }
+{\rm H. c.} \right]
\nonumber
\\
&+&\int d^3 r d^3 r' 
{ {|\Delta({\bf r}, {\bf r}')|^2} \over {V_{\rm eff}({\bf r}, {\bf r}') }}
\end{eqnarray}
where the gap function $\Delta({\bf r}, {\bf r}')$ is defined as 
\begin{eqnarray}
 \Delta({\bf r}, {\bf r}')= V_{\rm eff}({\bf r}, {\bf r}')\langle e^{{i \over 2}(\hat{\chi}({\bf r}) +\hat{\chi}({\bf r}')) }
\hat{\Psi}_{\uparrow}({\bf r}) \hat{\Psi}_{\downarrow} ({\bf r'}) \rangle
\end{eqnarray}

Due to the factor $ e^{{i \over 2}(\hat{\chi}({\bf r}) +\hat{\chi}({\bf r}')}$ the expectation value can be calculated using the particle number fixed state.

Using commutation relations for $\hat{\Psi}^{\dagger}_{\sigma }({\bf r})$ and $\hat{\Psi}_{\sigma }({\bf r})$,
\begin{eqnarray}
&&\{ \hat{\Psi}_{\sigma }({\bf r}),\hat{\Psi}^{\dagger}_{\sigma' }({\bf r}') \}=\delta_{\sigma \sigma'}\delta({\bf r} -{\bf r}')
\nonumber
\\
&&\{ \hat{\Psi}_{\sigma }({\bf r}),\hat{\Psi}_{\sigma' }({\bf r}') \}=0
\nonumber
\\
&&\{ \hat{\Psi}^{\dagger}_{\sigma }({\bf r}),\hat{\Psi}^{\dagger}_{\sigma' }({\bf r}') \}=0
 \end{eqnarray}
the following relations are obtained
\begin{eqnarray}
\left[\hat{\Psi}_{\uparrow }({\bf r}) , {H}_{\rm MF} \right]&=&
{h}({\bf r})\hat{\Psi}_{\uparrow }({\bf r})+\int d^3 r' \Delta({\bf r},{\bf r}')\hat{\Psi}^{\dagger}_{\downarrow }({\bf r}')e^{-{i \over 2}(\hat{\chi}({\bf r}) +\hat{\chi}({\bf r}')) }
\nonumber
\\
\left[\hat{\Psi}_{\downarrow }({\bf r}) , {H}_{\rm MF} \right] &=&{h}({\bf r})\hat{\Psi}_{\downarrow }({\bf r})-\int d^3 r' \Delta({\bf r},{\bf r}')\hat{\Psi}^{\dagger}_{\uparrow }({\bf r}')e^{-{i \over 2}(\hat{\chi}({\bf r}) +\hat{\chi}({\bf r}')) }
\label{deG1}
\end{eqnarray}

The particle number conserving Bogoliubov operators $\gamma_{n \sigma}$ and $\gamma^{\dagger}_{n \sigma}$ obey fermion commutation relations. They are chosen to satisfy
\begin{eqnarray}
\left[ {H}_{\rm MF}, \gamma_{n \sigma } \right]=-\epsilon_n \gamma_{n \sigma}, \quad \left[{H}_{\rm MF}, \gamma^{\dagger}_{n \sigma } \right] =\epsilon_n \gamma^{\dagger}_{n \sigma}
\label{deG2}
\end{eqnarray}
yielding the diagonalized ${H}_{\rm MF}$,
\begin{eqnarray}
{H}_{\rm MF}=E_g + \sum_{n, \sigma} \epsilon_n \gamma^{\dagger}_{n \sigma}\gamma_{n \sigma}
\label{deG3}
\end{eqnarray}
where the excitation energies satisfy $\epsilon_n > 0$, and $E_g$ is the ground state energy.

From Eqs.~(\ref{deG1}), (\ref{deG2}), and (\ref{deG3}), the following system of equations are obtained
\begin{eqnarray}
\epsilon_n e^{-{i \over 2} \hat{\chi}({\bf r})}u_n({\bf r})&=&
h({\bf r}) e^{-{i \over 2}\hat{\chi}({\bf r})}u_n({\bf r})+\int d^3 r' \Delta ({\bf r},{\bf r}')e^{-{i \over 2}\hat{\chi}({\bf r})}v_n({\bf r}')
\nonumber
\\
\epsilon_n e^{-{i \over 2} \hat{\chi}({\bf r})}v^{\ast}_n({\bf r})&=&-
 h({\bf r}) e^{-{i \over 2} \hat{\chi}({\bf r})}v^{\ast}_n({\bf r})+\int d^3r' \Delta ({\bf r},{\bf r}')e^{-{i \over 2}\hat{\chi}({\bf r})}u^{\ast}_n({\bf r}')
\end{eqnarray}

Using the relation in Eq.~(\ref{eqBC}),
the above are cast into the following,
\begin{eqnarray}
\epsilon_n u_n({\bf r})&=&
\bar{h}({\bf r}) u_n({\bf r})+\int d^3 r'\Delta ({\bf r},{\bf r}')v_n({\bf r}')
\nonumber
\\
\epsilon_n v_n({\bf r})&=&-
 \bar{h}^{\ast}({\bf r}) v_n({\bf r})+\int d^3 r'\Delta^{\ast}({\bf r},{\bf r}')u_n({\bf r}')
 \label{e1}
\end{eqnarray}
where 
\begin{eqnarray}
\bar{h}({\bf r})={ 1 \over {2m_e}} \left( { \hbar \over i} \nabla +{e \over c} {\bf A}^{\rm em}-{ \hbar \over 2} \nabla \chi \right)^2+U({\bf r})-\mu 
 \label{e2}
\end{eqnarray}
and 
\begin{eqnarray}
\Delta({\bf r}, {\bf r}')=V_{\rm eff}({\bf r}, {\bf r}')\sum_n \left[ u_n({\bf r}) v^{\ast}_n({\bf r}')(1- f(\epsilon_n))-u_n({\bf r}') v^{\ast}_n({\bf r})f(\epsilon_n) \right]
 \label{e3}
\end{eqnarray}
$f(\epsilon_n)$ is the Fermi function given by
\begin{eqnarray}
f(\epsilon_n)={1 \over {e^{\epsilon_n \over {k_{\rm B} T}}+1}}
\end{eqnarray}
where $k_{\rm B}$ is the Boltzmann constant.

The above system of equations are Bogoliubov-de~Gennes equations \cite{deGennes} 
using the particle number conserving Bogoliubov operators \cite{koizumi2019}.

Note that the gauge potential in the single particle Hamiltonian $\bar{h}({\bf r})$ is the effective one given by
\begin{eqnarray}
{\bf A}^{\rm eff}={\bf A}^{\rm em}-{ {\hbar c} \over {2e}} \nabla \chi
\end{eqnarray}

In the BCS superconductor, the energy gain by the electron-pair formation exceeds the energy loss by performing the spin-twisting itinerant motion that is necessary to have the number changing operators; thus, even if the spin-orbit interaction is negligibly small, the spin-twisting itinerant motion will occur. This will be the reason for the simultaneous occurrence of superconductivity and energy gap formation.
 
By solving the system of equations composed of Eqs.~(\ref{e1}), (\ref{e2}), and (\ref{e3}), with the condition ${\bf A}^{\rm eff}=0$, we obtain the currentless solutions
for $u_n, v_n$, which we denote as $\acute{u}_n, \acute{v}_n$. 

Using $\acute{u}_n, \acute{v}_n$, we express  ${u}_n, {v}_n$ as
\begin{eqnarray}
u_n({\bf r})=\acute{u}_n ({\bf r}) e^{{i \over 2}{\chi}({\bf r})}, \quad v_n({\bf r})=\acute{v}_n ({\bf r}) e^{-{i \over 2}{\chi}({\bf r})}
\label{equv}
\end{eqnarray}

We obtain $\chi$ in the following manner.
First, we note that $\chi$ has to satisfy constraints,
\begin{eqnarray}
 \oint_{C_\ell} \nabla \chi \cdot d {\bf r}=2 \pi w_{C_{\ell}}[\chi]
 \label{eqConstraint}
\end{eqnarray}
where $w_{C_{\ell}}[\chi]$ is the integer called the winding number. The value $w_{C_{\ell}}[\chi]$ needs to satisfy the condition similar to the one in Eq.~(\ref{wcond}) to make the wave function single-valued.

We obtain $\chi$ for the ground state by minimizing the total energy under the above constraint.
For this purpose, we use the following functional,
\begin{eqnarray}
F[\nabla \chi]=E[\nabla \chi]+\sum_{\ell=1}^{N_{\rm loop}} { {\lambda_{\ell}}}\left(  \oint_{C_\ell} \nabla \chi \cdot d {\bf r}-2 \pi w_{C_{\ell}}[\chi] \right), 
\label{functional}
\end{eqnarray}
where 
\begin{eqnarray}
E[\nabla \chi]=\langle {\rm Gnd}| H_{\rm eff} |{\rm Gnd} \rangle
\label{energyf}
\end{eqnarray}
is obtained using Eq.~(\ref{equv});
 $\lambda_{\ell}$'s are Lagrange multipliers, and $\{ C_1, \cdots, C_{N_{\rm loop}} \}$ are boundaries of plaques of the lattice (we assume we have the lattice version of the Hamiltonian); and $N_{\rm loop}$ is the total 
number of plaques of the lattice. 

The minimum condition of  $F[\nabla \chi]$ with respect to the variation of $\nabla \chi$ yields,
\begin{eqnarray}
{{\delta E[\nabla \chi]} \over {\delta \nabla \chi}}+\sum_{\ell=1}^{N_{\rm loop}} { {\lambda_{\ell}}} {{\delta } \over {\delta \nabla \chi}} \oint_{C_\ell} \nabla \chi \cdot d {\bf r}=0
 \label{eqMinimum}
\end{eqnarray}

Then, the current density is given by
\begin{eqnarray}
{\bf J}={{2e} \over \hbar}  {{\delta E} \over {\delta \nabla \chi}}=-{{2e} \over \hbar} 
\sum_{\ell=1}^{N_{\rm loop}} { {\lambda_{\ell}}} {{\delta } \over {\delta \nabla \chi}} \oint_{C_\ell} \nabla \chi \cdot d {\bf r}
\label{loopC}
\end{eqnarray}
This indicates that the current flowing in superconducting states is a collection of loop currents.

Expressing  Eqs.~(\ref{eqConstraint}) and (\ref{eqMinimum}) using the discrete lattice Hamiltonian, a system of equations for $\nabla \chi$ is obtained,
\begin{eqnarray}
 \sum_{k \leftarrow j} L_{k \leftarrow j}^{\ell}\tau_{k \leftarrow j} &=& 2 \pi w_{C_{\ell}}[\chi]
 \label{Feq2b}
 \\
{{\partial  E( \{ \tau_{k \leftarrow j} \}) } \over {\partial \tau_{k \leftarrow j} }}&+&\sum_{\ell=1}^{N_{\rm loop}}  { {\lambda_{\ell}}} 
{{\partial } \over {\partial \tau_{k \leftarrow j}}} \sum_{k \leftarrow j} L_{k \leftarrow j}^{\ell}\tau_{k \leftarrow j} =0
\label{Feq1b}
\end{eqnarray}
where ${k \leftarrow j}$ indicates the bond that starts from site $j$ and ends at site $k$, $\tau_{ k \leftarrow j}$ is the difference of $\chi$ for the bond $k \leftarrow j $
\begin{eqnarray}
\tau_{ k \leftarrow j}=\chi_k -\chi_j 
\end{eqnarray}
and $L_{k \leftarrow j}^{\ell}$ is defined as
\begin{eqnarray}
L_{k \leftarrow j}^{\ell} =
\left\{
\begin{array}{cl}
-1 & \mbox{ if  $ k \leftarrow j$ exists in $C_{\ell}$ in the clockwise direction}
\\
1 & \mbox{ if  $ k \leftarrow j$ exists in $C_{\ell}$ in the counterclockwise direction}
\\
0 & \mbox{ if  $ k \leftarrow j$ does not exist in $C_{\ell}$}
\end{array}
\right.
\end{eqnarray}
 Note that a set of parameters $\{ w_{C_{\ell}}[\chi] \}$ must be supplied as part of boundary conditions.
 
We take the branch of $\chi_j$ that satisfies the difference of value from the nearest neighbor site $k$ is in the range,
\begin{eqnarray}
-\pi \le \chi_{j} -\chi_k < \pi
\end{eqnarray}

\begin{figure}
\begin{center}
\includegraphics[width=8.0cm]{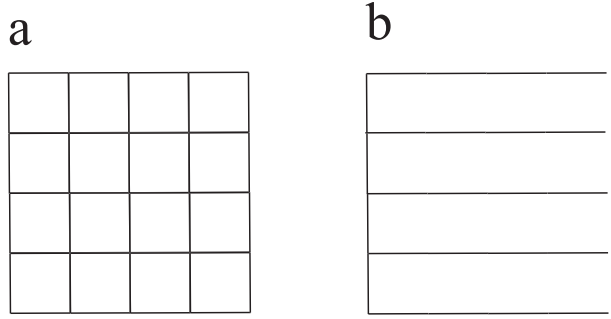}
\end{center}
\caption{An example of a simply-connected lattice constructed from square lattice. $5 \times 5$ square lattice in ${\bf a}$ becomes a simply-connected lattice in ${\bf b}$ by removing 16 bonds.}
\label{LatticeSimple}
\end{figure}

From ($\chi_j-\chi_k$)'s, we rebuild $\chi$. 
The rebuilding process is as follows: first, we pick a value for the initial $\chi_1$ (say $\chi_1=0$). 
After fixing the value of $\chi_1$, we calculate $\chi_2$ by $\chi_{2} = \chi_1 + (\chi_{2} -\chi_1)$, where the site $2$ is connected to the site $1$ by a nearest neighbor bond.
 The step where value $\chi_{j}$ is derived from the already evaluated value of $\chi_k$ is given by
\begin{eqnarray}
\chi_{j} = \chi_k + (\chi_{j} -\chi_k)
\end{eqnarray}
where the sites ${j}$ and $k$ are connected by a bond in the path for the rebuilding of $\chi$. 
This process is continued until values at all accessible sites are evaluated once and only once. 

By this rebuilding process, a single path is constructed from the site $1$ to other sites $k\neq 1$, which is achieved by making the region simply-connected by removing some bonds (see Fig.~\ref{LatticeSimple}). 
We denote the path from the site $1$ to other sites $k\neq 1$ by $C_{1 \rightarrow k}$.
Then, the value $\chi_k$ is given by
\begin{eqnarray}
\chi_k \approx \chi_1+ \int_{C_{1 \rightarrow k}} \nabla \chi \cdot d{\bf r}
\end{eqnarray}

When a magnetic field ${\bf B}^{\rm em}=\nabla \times {\bf A}^{\rm em}$ is applied,
 the energy functional  in Eq.~(\ref{energyf}) is modified as
\begin{eqnarray}
E[\nabla \chi] \  \rightarrow \ E \left[\nabla \chi -{{2e} \over {\hbar c}}{\bf A}^{\rm em} \right]
\label{energyfAem}
\end{eqnarray}

This leads to replace $\tau_{ j \leftarrow i}$ in $E$ by
\begin{eqnarray}
\tau_{ j \leftarrow i} \ \rightarrow  \ \tau_{ j \leftarrow i}-{{2e} \over {\hbar c}} \int^j_i{\bf A}^{\rm em}\cdot d{\bf r}
\end{eqnarray}
By this modification, a magnetic field can be taken into account.

Lastly in this section, we would like to consider the mass in the London moment.
 In the present formalism, the velocity field ${\bf v}$ is given by
\begin{eqnarray}
{\bf v}={ e\over {m_e c}}{\bf A}^{\rm em}-{\hbar \over {2 m_e}} \nabla \chi
\end{eqnarray}
 since the velocity field from $\tilde{u}_n, \tilde{v}_n$ is zero. 
 This indicates that $m$ in Eq.~(\ref{Londoneq4b}) is $m_e$ in agreement with the value observed in the London moment.

\section{Model calculations}
 \label{sec5}

\subsection{Model Hamiltonian}

In this section, a model is solved using the particle number conserving version of the Bogoliubov-de Gennes equations.
This model is originally built to study the cuprate superconductivity. 
Since the purpose of the present work is to explain the new theory, the model calculation below should be consider as an example exercise. 
The validity of the model as a model for the cuprate superconductivity will be dealt elsewhere.

\begin{figure}
\begin{center}
\includegraphics[width=12.0cm]{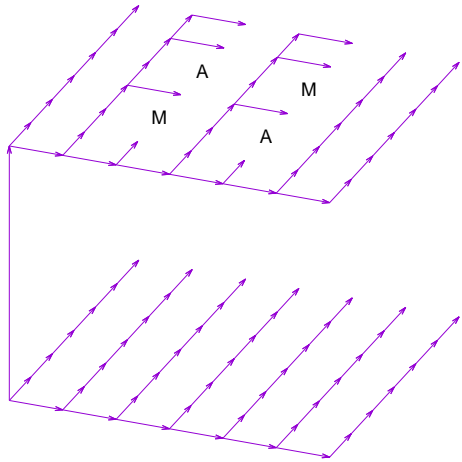}
\end{center}
\caption{Simply-connected lattice constructed by removing some bonds for the two-layer model. We consider the model with two 7$\times$7 square lattices stacked in the $z$ direction.
There are four small polarons in the bulk layer indicated by ``M'' and ``A''.  Each ``M'' indicates a center of a spin-vortex with winding number $+1$,
and ``A'' a center of a spin-vortex with winding number $-1$.}
\label{Lattice2}
\end{figure}

The model is composed of two layers, a surface layer and
a bulk layer (see Fig.~\ref{Lattice2}). The superconductivity occurs in the bulk, however, many experimental results show contributions from the surface region, thus, the surface layer effects must be included to interpret experiments.
The experimental results seem to indicate that the small polaron formation and coexisting magnetic moment effects are strong in the bulk, however, they are weak in the surface region.

Thus, we developed a model containing two layers to reproduce experimental results.
It is also notable that although the electron-pairing does not occur in the bulk layer, supercurrent flows there in this model;
thus, it demonstrates that the primary importance for superconductivity is the appearance of the non-trivial Berry connection.

\subsection{Bulk layer}

The model for the bulk layer is constructed based on the fact that the parent compound is a Mott insulator \cite{AndersonBook}, which is
well-described by the two dimensional Hubbard model with large on-site repulsion
\begin{eqnarray}
U \gg t_1
\end{eqnarray}
where $U$ is the on-site repulsion parameter and $t_1$ is the nearest neighbor hopping parameter.
The bulk CuO$_2$ plane of the cuprate is taken as a two-dimensional square lattice, where a copper atom resides at each site.
Oxygens between copper atoms are not explicitly taken into account in the present model.

When holes are doped, they form small polarons in the bulk \cite{Bianconi,Miyaki2008}. At low temperatures, the mobility of the small polaron is very small; thus, the supercurrent is not due to the hopping of small polarons.

The inelastic neutron scattering experiments indicate spin-wave excitations with spin lying in CuO$_2$ plane \cite{Kastner1998}. Thus, the magnetic moment coexisting with superconductivity in the bulk. The magnetic excitation spectrum has a hourglass-shaped dispersion, suggesting the stripes of spins or spin-vortices \cite{Neutron,Hidekata2011}.

By taking into account the above observation, the following bulk Hamiltonian is constructed,
\begin{eqnarray}
     H_{\rm{EHFS}} &=& 
     -t_1\sum_{\langle i,j \rangle_{1},\sigma}[c^\dagger_{i\sigma}c_{j\sigma}+{\rm H.c.}]
   +U \sum_j c^\dagger_{j\uparrow} c_{j\uparrow} c^\dagger_{j\downarrow} c_{j\downarrow} 
     +J_h \sum_{\left<i,j\right>_h}{\bf S}_i \cdot {\bf S}_j
    \nonumber
    \\
    &+& \lambda\sum_{h} [
    c^{\dagger}_{h+y \downarrow} c_{h-x \uparrow}
    -c^{\dagger}_{h+y \uparrow}c_{h-x \downarrow}
    + c^{\dagger}_{h+x \downarrow} c_{h-y \uparrow}
    - c^{\dagger}_{h+x \uparrow} c_{h-y \downarrow} 
    \nonumber
    \\
    &+& \text{i}(c^{\dagger}_{h-x\downarrow}c_{h-y\uparrow}
    + c^{\dagger}_{h-x\uparrow}c_{h-y\downarrow})
    + \text{i}(c^{\dagger}_{h+y\downarrow}c_{h+x\uparrow}
    + c^{\dagger}_{h+y\uparrow}c_{h+x\downarrow})  
    + \text{h.c.}])
    \nonumber
    \\
    &-&\mu_{\rm bulk} \sum_{j, \sigma} c_{j \sigma}^{\dagger}c_{j \sigma}
  \end{eqnarray}
  where $c^{\dagger}_{j \sigma}$ and  $c_{j \sigma}$ are creation and annihilation operators for the electron at $j$th site with spin $\sigma$, respectively.
  
   In this Hamiltonian, the doped holes are assumed to form small polarons and immobile, thus, the effectively-half-filled situation (EHFS) is realized, where the number of electrons and the number of accessible sites for the electron are equal.
  We take the hole-occupied sites as inaccessible sites for electrons, thus, not included in the sum.
  
  The first term describes the electron hopping with transfer integral $t_1$; $\langle i,j\rangle_1$ indicates that the sum is taken over
  1st nearest neighbor hoppings.  
   
  The second term describes the on-site Coulomb repulsion; $U$ is significantly larger that $t_1$ (actually, we adopt $U=8t_1$ in the present work). 
  
  The third term describes the antiferromagnetic exchange interaction between electron spins around each doped hole; $\langle i,j\rangle_h$ indicates that the sum is taken over pairs around holes, i.e., each hole (denoted by $h$) accompanies four sites around it (denoted by $h-x, h+x, h-y$ and $h+y$;
  they are the nearest neighbor sites of $h$ in $-x$, $+x$, $-y$, and $+y$ directions, respectively), and there are six site-pairs that contribute to $\langle i,j\rangle_h$ for each doped hole. The parameter $J_h$ is taken to be $0.5J_{\rm AF}$ where $J_{\rm AF}={{4 t_1^2 } \over U}$ is the antiferromagnetic exchange parameter for the parent compound \cite{HKoizumi2015B}.
  
  The fourth term is the Rashba interaction term \cite{Rashba2013}, which we only include around the holes \cite{Koizumi2017}. The fifth term is the chemical potential term.
  
 Components of the spin operator ${\bf S}_j=({S}^{x}_j, S^{y}_j, S^{z}_j)$ are given by
  \begin{eqnarray}
 {S}^{x}_j&=&\frac{1}{2}( c^{\dagger}_{j\uparrow} c_{j\downarrow}+c^{\dagger}_{j\downarrow} c_{j\uparrow})
 \nonumber
 \\
 {S}^{y}_j&=&\frac{i}{2}(-c^{\dagger}_{j\uparrow}\ c_{j\downarrow}+c^{\dagger}_{j\downarrow} c_{j\uparrow})
 \nonumber
 \\
 {S}^{z}_j&=&\frac{1}{2}( c^{\dagger}_{j\uparrow} c_{j\uparrow}-c^{\dagger}_{j\downarrow} c_{j\downarrow})
\end{eqnarray}

The chemical potential $\mu_{\rm bulk}$ is chosen to control the number of small polarons in EHFS.

\begin{figure}
\begin{center}
\includegraphics[width=10.0cm]{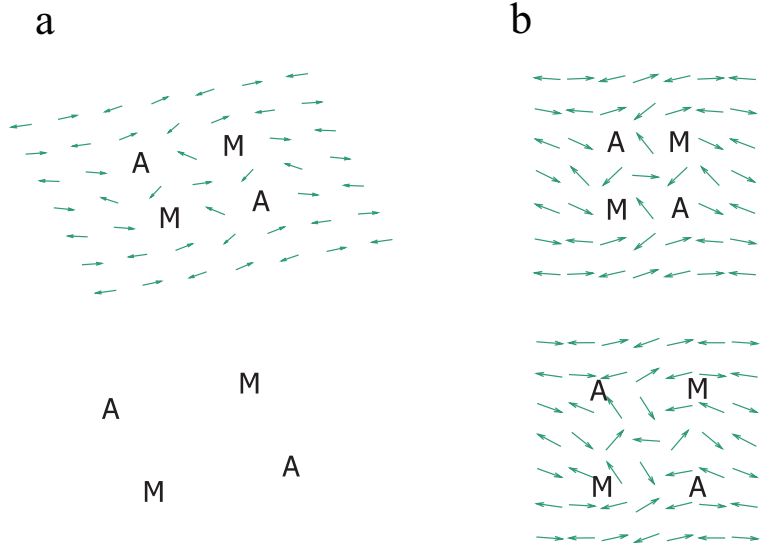}
\end{center}
\caption{Spin texture obtained. {\bf a}: The spin-texture of the two layer. The spins in the surface layer are very small.
{\bf b}: The normalized spin-textures in the bulk (top), and that in the surface (bottom). Spin-vortices are created in the surface layer due to the influence of those
in the bulk layer.}
\label{Spins}
\end{figure}

Since $H_{\rm{EHFS}}$ is too difficult to handle as it is, we use a mean field version of it
 \begin{eqnarray}
     H^{\rm{HF}}_{\rm{EHFS}} &=& 
     -t_1\sum_{\langle i,j \rangle_{1},\sigma}[c^\dagger_{i\sigma}c_{j\sigma}+{\rm H.c.}]
   \nonumber
   \\
     &+& U \sum_j [(-\frac{2}{3}\langle {S^z_j} \rangle + \frac{1}{2})c^\dagger_{j\uparrow} c_{j\uparrow}
     +(\frac{2}{3}\langle{S^z_j} \rangle+ \frac{1}{2}) c^\dagger_{j\downarrow} c_{j\downarrow} 
     \nonumber
     \\
     &-&\frac{2}{3}(\langle{S^x_j}\rangle - \rm{i} \langle{S^y_j}\rangle) c^\dagger_{j\uparrow} c_{j\downarrow}
       -\frac{2}{3}(\langle{S^x_j} \rangle+ \rm{i} \langle{S^y_j}\rangle) c^\dagger_{j\downarrow} c_{j\uparrow}
     - \frac{2}{3} \langle{\bf S}_j\rangle^2 ]
     \nonumber
     \\
    &+& J_h \sum_{\left<i,j\right>_h}
    [\frac{1}{2}(\langle{S^z_i} \rangle c^\dagger_{j\uparrow} c_{j\uparrow} + \langle{S^z_j}\rangle  c^\dagger_{i\uparrow} c_{i\uparrow})
    - \frac{1}{2}(\langle{S^z_i}\rangle c^\dagger_{j\downarrow} c_{j\downarrow} + \langle{S^z_j} \rangle c^\dagger_{i\downarrow} c_{i\downarrow})
    \nonumber
     \\
    & +& \frac{1}{2}\{(\langle{S^x_i}\rangle - \rm{i} \langle{S^y_i} \rangle ) c^\dagger_{j\uparrow} c_{j\downarrow} + (\langle{S^x_j} \rangle- \rm{i} \langle{S^y_j}\rangle) c^\dagger_{i\uparrow} c_{i\downarrow}\}   
    \nonumber
    \\
    &+&\frac{1}{2}\{(\langle{S^x_i}\rangle + \rm{i} \langle{S^y_i}\rangle) c^\dagger_{j\downarrow} c_{j\uparrow} + (\langle{S^x_j}\rangle + \rm{i} \langle{S^y_j}\rangle) c^\dagger_{i\downarrow} c_{i\uparrow}\}
    - \langle{\bf S}_i\rangle \cdot \langle {\bf S}_j \rangle] 
    \nonumber
    \\
    &+& \lambda\sum_{h} [
    c^{\dagger}_{h+y \downarrow} c_{h-x \uparrow}
    -c^{\dagger}_{h+y \uparrow}c_{h-x \downarrow}
    + c^{\dagger}_{h+x \downarrow} c_{h-y \uparrow}
    - c^{\dagger}_{h+x \uparrow} c_{h-y \downarrow} 
    \nonumber
    \\
    &+& \text{i}(c^{\dagger}_{h-x\downarrow}c_{h-y\uparrow}
    + c^{\dagger}_{h-x\uparrow}c_{h-y\downarrow})
    + \text{i}(c^{\dagger}_{h+y\downarrow}c_{h+x\uparrow}
    + c^{\dagger}_{h+y\uparrow}c_{h+x\downarrow})  
    + \text{h.c.}]
    \nonumber
    \\
    &-&\mu_{\rm bulk} \sum_{j, \sigma} c_{j \sigma}^{\dagger}c_{j \sigma}
    \label{SV2:H^HF_EHFS-2}
  \end{eqnarray}
  where $\langle \hat{O} \rangle$ denotes the expectation value of the operator $\hat{O}$. 
  This Hamiltonian yields states with spin-vortices around small polarons (see Fig.\ref{Spins}).
  These spin-vortices induce loop currents called the ``{\em spin-vortex-induced loop currents}'', due to the appearance of a non-trivial Berry connection.
  
 \subsection{Surface layer}
 
The Hamiltonian for the surface layer is following
 \begin{eqnarray}
 H_{\rm surf}
  &=&-t_1\sum_{\langle i,j \rangle_{1},\sigma}[c^\dagger_{i\sigma}c_{j\sigma}+{\rm H.c.}]
   -t_2 \sum_{\langle i,j \rangle_{2},\sigma}[c^\dagger_{i\sigma}c_{j\sigma}+{\rm H.c.}]
    +U \sum_j c^\dagger_{j\uparrow} c_{j\uparrow} c^\dagger_{j\downarrow} c_{j\downarrow} 
-\mu_{\rm surf} \sum_{j, \sigma} c_{j \sigma}^{\dagger}c_{j \sigma}
\nonumber
\\
  \end{eqnarray}
  This is the Hubbard model including the second nearest neighbor hopping.
  The second nearest neighbor hopping with the parameter $t_2$ is included to have a curved Fermi surface in the angle-resolved photoemission spectrum. We use $t_2=-0.12t_1$ in this work.
  
  The most important difference from the bulk Hamiltonian is the absence of the small polaron effect. The spin-orbit interaction term is also absent.
  The chemical potential $\mu_{\rm surf}$ is so chosen that 
  the bulk and surface hole densities are equal.
  
   Since $H_{\rm surf}$ is too difficult to handle as it is, we construct a mean-field version via the $t-J$ model \cite{Zhang1988} for the Hubbard model,
\begin{eqnarray}
 H^{\rm HF}_{\rm surf}
  &=&-t_1\sum_{\langle i,j \rangle_{1},\sigma}[(1-\langle n_{i, -\sigma} \rangle)c^\dagger_{i\sigma}c_{j\sigma}(1-\langle n_{j, -\sigma} \rangle)+{\rm H.c.}]
  \nonumber
  \\
   &-&t_2 \sum_{\langle i,j \rangle_{2},\sigma}[(1-\langle n_{i, -\sigma} \rangle)c^\dagger_{i\sigma}c_{j\sigma}(1-\langle n_{j, -\sigma} \rangle)+{\rm H.c.}]
         \nonumber
   \\
     &+&\sum_{\langle i,j \rangle_{1}}(\Delta_{ij} e^{ -{i \over 2} \hat{\chi}_j} e^{ -{i \over 2} \hat{\chi}_i} c_{i \uparrow}^{\dagger}c_{j \downarrow}^{\dagger} + {\rm H.c.})-\mu_{\rm surf}  \sum_{j, \sigma} c_{j \sigma}^{\dagger}c_{j \sigma}
  \end{eqnarray}
  where the pairing amplitude or pair potential $\Delta_{ij}$ is defined as 
  \begin{eqnarray}
  \Delta_{ij}=-{{4t_1^2} \over U} \langle e^{ {i \over 2} \hat{\chi}_j}e^{ {i \over 2} \hat{\chi}_i}(c_{j \downarrow}c_{i \uparrow}
     -c_{j \uparrow}c_{i \downarrow} )\rangle= \Delta_{ji}
  \end{eqnarray}
  Due to the presence of the number changing operators, $e^{ {i \over 2} \hat{\chi}_j}$ and $e^{ {i \over 2} \hat{\chi}_i}$,
   $\Delta_{ij}$ is obtained in the particle number conserving formalism. The appearance of the number changing operator 
   is due to the spin-vortex formation in the bulk; thus, the electron pair formation is a secondary effect.
   To have the superconducting state, stable loop currents that can generate a macroscopic persistent current should be realized.
   
   The pair potential $\Delta_{ij}$ gives rise to $d$-wave pairing gap, and explains the Fermi arc \cite{ArpesRev}. 
   We would like to emphasized that the appearance of the $d$-wave pairing gap
   is secondary effect enabled by the appearance of the non-trivial Berry connection in the bulk; this generates the collective mode for supercurrent,
   and particle number changing operators. 
   The particle number changing operators make it possible to gain energy by the $d$-wave pairing gap formation.

   \subsection{Interlayer hopping}
   
   The bulk and surface layers are connected by the inter-layer hopping.
   The following is the inter-layer hopping Hamiltonian,
   \begin{eqnarray}
 H_{\rm inter-layer}
  &=&-t_3\sum_{\langle i_s,j_b \rangle_{z}, \sigma}[(1-n_{i_s, -\sigma})c^\dagger_{i_s\sigma}c_{j_b\sigma}(1-n_{j_b, -\sigma})+{\rm H.c.}+{{4t_3^2} \over U} \sum_{\langle i_s,j_b \rangle_{z}}\left( {\bf S}_{i_s}\cdot {\bf S}_{j_b} -{1 \over 4}n_{i_s} n_{j_b} \right)
\nonumber
\\
  \end{eqnarray}
  where an antiferromagnetic exchange interaction between spins in the bulk-layer and those in the surface-layer is included using a $t-J$ model approximation.
  
Since  $H_{\rm inter-layer}$ is still too difficult to handle, we use the following mean field version,
  \begin{eqnarray}
   &&H^{\rm HF}_{\rm inter-layer}
   =
  -t_3\sum_{\langle i_s,j_b \rangle_{z}, \sigma}[(1-\langle n_{i_s, -\sigma} \rangle )c^\dagger_{i_s\sigma}c_{j_b\sigma}(1-\langle n_{j_b, -\sigma} \rangle)+{\rm H.c.}]
         \nonumber
       \\
  &+&{{2t_3^2} \over U} \sum_{\langle i_s,j_b \rangle_{z}}\Big[ ( \langle { S}^x_{i_s} \rangle -i   \langle { S}^y_{i_s} \rangle) c^{\dagger}_{j_b \uparrow} c_{j_b \downarrow}+ (\langle { S}^x_{i_s} \rangle + i   \langle { S}^y_{i_s} \rangle) c^{\dagger}_{j_b \uparrow} c_{j_b \downarrow}
  \nonumber
  \\
  &+&
 ( \langle { S}^z_{i_s} \rangle-{1 \over 2} \langle n_{i_s} \rangle )c^{\dagger}_{j_b \uparrow} c_{j_b \uparrow}- ( \langle { S}^z_{i_s} \rangle+{1 \over 2} \langle n_{i_s} \rangle )c^{\dagger}_{j_b \downarrow} c_{j_b \downarrow}) \Big]
  \nonumber
  \\
    &+&{{2t_3^2} \over U} \sum_{\langle i_s,j_b \rangle_{z}}\Big[ ( \langle { S}^x_{j_b} \rangle -i   \langle { S}^y_{j_b} \rangle) c^{\dagger}_{i_s \uparrow} c_{i_s \downarrow}+ (\langle { S}^x_{j_b} \rangle + i   \langle { S}^y_{j_b} \rangle) c^{\dagger}_{i_s \uparrow} c_{i_s \downarrow}
    \nonumber
    \\
    &+&
   ( \langle { S}^z_{j_b} \rangle-{1 \over 2} \langle n_{j_b} \rangle )c^{\dagger}_{i_s \uparrow} c_{i_s \uparrow}
   -( \langle { S}^z_{j_b} \rangle+{1 \over 2} \langle n_{j_b} \rangle)  c^{\dagger}_{i_s \downarrow} c_{i_s \downarrow}) \Big]
  \nonumber
  \\
  &-&{{4t_3^2} \over U} \sum_{\langle i_s,j_b \rangle_{z}}\left( \langle {\bf S}_{i_s}\rangle \cdot \langle {\bf S}_{j_b}\rangle -{1 \over 4}\langle n_{i_s}\rangle \langle n_{j_b} \rangle \right)
  \end{eqnarray}
  
  We adopt $t_3=0.01t_1$ in the following calculations.
  Through this Hamiltonian, the non-trivial Berry connection generated in the bulk propagates in the surface.
  
    \subsection{The total Hamiltonian for the two-layer model}
  
  Over all, our model Hamiltonian is given by
  \begin{eqnarray}
H_{\rm eff}
&=&H^{\rm HF}_{\rm EHFS}+H^{\rm HF}_{\rm surf}+H^{\rm HF}_{\rm inter-layer}
\nonumber
\\
&=&
\sum_{i,j,\sigma, \sigma'} h_{i \sigma, j \sigma'} c_{i \sigma}^{\dagger} c_{j \sigma'}
  + \sum_{i,j} [ \Delta_{ij} e^{ -{i \over 2} \hat{\chi}_j}e^{ -{i \over 2} \hat{\chi}_i}c_{i \uparrow}^{\dagger}c_{j \downarrow}^{\dagger}+ \Delta^{\ast}_{ij} e^{ {i \over 2} \hat{\chi}_j}e^{ {i \over 2} \hat{\chi}_i}c_{j \downarrow}c_{i \uparrow}]
  +E_{\rm const}
  \nonumber
  \\
  \label{Heff}
  \end{eqnarray}

 \subsection{Particle number conserving Bogoliubov-de Gennes equations}
 
 In order to obtain the particle number conserving Bogoliubov-de Gennes equations from $H_{\rm eff}$ in Eq.~(\ref{Heff}), we perform the following Bogoliubov transformation
  \begin{eqnarray}
 c_{ i \uparrow} &=&\sum'_{n} [ u^{n}_{i \uparrow}\gamma_{n}- (v^{n}_{i \uparrow})^{\ast}\gamma_{n}^{\dagger}] e^{ -{i \over 2} \hat{\chi}_i}
 \nonumber
 \\
 c_{ i \downarrow} &=&\sum'_{n} [ u^{n}_{i \downarrow}\gamma_{n}+ (v^{n}_{i \downarrow})^{\ast}\gamma_{n}^{\dagger}] e^{ -{i \over 2} \hat{\chi}_i}
 \label{eq137}
  \end{eqnarray}
  
 Using the Bogoliubov operators $\gamma_n$ and $\gamma^{\dagger}_n$, 
 $H_{\rm eff}$ is expressed as
    \begin{eqnarray}
H_{\rm eff}=\sum_{n}'  E_n  \gamma^{\dagger}_{n} \gamma_{n} + E'_{\rm const.}
  \end{eqnarray}
  where $E'_{\rm const.}$ is a constant and ``$ \sum_n' $'' denotes that the sum is take over $E_n >0$,
  and the ground state $|{\rm Gnd} \rangle$ satisfies
      \begin{eqnarray}
    \gamma_n |{\rm Gnd} \rangle =0, \quad E_n >0
    \end{eqnarray}
  
  The Bogoliubov operators are fermion operators that satisfy
  \begin{eqnarray}
  \{  \gamma_{n}, \gamma^{\dagger}_{m} \}= \delta_{nm},  \quad \{  \gamma^{\dagger}_{n}, \gamma^{\dagger}_{m} \}=0, \quad  \{  \gamma_{n}, \gamma_{m} \}=0
  \end{eqnarray}
  thus, the commutation relations
        \begin{eqnarray}
[  \gamma^{\dagger}_{n}, H_{\rm eff}]=-E_n  \gamma^{\dagger}_{n}, \quad  [  \gamma_{n}, H_{\rm eff}]=E_n  \gamma_{n}
  \label{eq167}
  \end{eqnarray}
  are obtained.

The following commutation relations are obtained between $H_{\rm eff}$ and $c_{i \sigma}$, $c^{\dagger}_{i \sigma}$,
\begin{eqnarray}
&&[c_{i \uparrow}, H_{\rm eff}] =\sum_{j,\sigma'} h_{i \uparrow, j \sigma'}  c_{j \sigma'}
  + \sum_{j}  \Delta_{ij} e^{ -{i \over 2} \hat{\chi}_j} e^{ -{i \over 2} \hat{\chi}_i} c_{j \downarrow}^{\dagger}
  \nonumber
  \\
  &&  [c_{i \uparrow}^{\dagger},
 H_{\rm eff} ] 
 =-
 \sum_{j,\sigma'} h_{j \sigma', i \uparrow}  c^{\dagger}_{j \sigma'}
  - \sum_{j} \Delta^{\ast}_{ij} e^{ {i \over 2} \hat{\chi}_j}e ^{ {i \over 2} \hat{\chi}_i}c_{j \downarrow}
  \nonumber
  \\
  &&[c_{i \downarrow}, H_{\rm eff}]=\sum_{j,\sigma'} h_{i \downarrow, j \sigma'}  c_{j \sigma'}
  - \sum_{j}  \Delta_{ji} e^{ -{i \over 2} \hat{\chi}_j} e^{ -{i \over 2} \hat{\chi}_i} c_{j \uparrow}^{\dagger}
  \nonumber
  \\
  &&  [c_{i \downarrow}^{\dagger},
 H_{\rm eff} ] 
 =-
 \sum_{j,\sigma'} h_{j \sigma', i \downarrow}  c^{\dagger}_{j \sigma'}
  +\sum_{j} \Delta^{\ast}_{ji} e^{ {i \over 2} \hat{\chi}_j}e ^{ {i \over 2} \hat{\chi}_i}c_{j \uparrow}
  \label{eq162}
  \end{eqnarray}

Using Eqs.~(\ref{eq162}), (\ref{eq137}), and (\ref{eq167}), the Bogoliubov-de Gennes equations are obtained,
    \begin{eqnarray}
 E_{n} u^{n}_{i \uparrow}&=& \sum_{j \sigma'}  e^{ {i \over 2} \hat{\chi}_i} h_{i \uparrow, j \sigma'} e^{- {i \over 2} \hat{\chi}_j} u^{n}_{j \sigma'} + 
 \sum_j \Delta_{ij} v^{n}_{j \downarrow} 
  \nonumber
 \\
  E_{n} u^{n}_{i \downarrow}&=& \sum_{j \sigma'}  e^{ {i \over 2} \hat{\chi}_i} h_{i \downarrow, j \sigma'} e^{- {i \over 2} \hat{\chi}_j} u^{n}_{j \sigma'} + \sum_j \Delta_{ji} v^{n}_{j \uparrow} 
  \nonumber
 \\
  E_{n} v^{n}_{i \uparrow}&=& -\sum_{j } e^{ -{i \over 2} \hat{\chi}_i} h^{\ast}_{i \uparrow, j \uparrow} e^{ {i \over 2} \hat{\chi}_j} v^{n}_{j \uparrow} +\sum_{j}  e^{ -{i \over 2} \hat{\chi}_i} h^{\ast}_{i \uparrow, j \downarrow} e^{ {i \over 2} \hat{\chi}_j} v^{n}_{j \downarrow} + \sum_j \Delta_{ij}^{\ast} u^{n}_{j \downarrow} 
  \nonumber
 \\
  E_{n} v^{n}_{i \downarrow}&=& \sum_{j} e^{ -{i \over 2} \hat{\chi}_i} h^{\ast}_{i \downarrow, j \uparrow} e^{ {i \over 2} \hat{\chi}_j} v^{n}_{j \uparrow} -\sum_{j} e^{ -{i \over 2} \hat{\chi}_i} h^{\ast}_{i \downarrow, j \downarrow} e^{ {i \over 2} \hat{\chi}_j} v^{n}_{j \downarrow} +\sum_j \Delta_{ji}^{\ast} u^{n}_{j \uparrow} 
  \end{eqnarray}
  
  In order to solve the above system of equations, we replace the operator $\hat{\chi}_j$ by its associated scalar value ${\chi}_j$. 
  Then, the Bogoliubov-de Gennes equations is put into the following matrix form,
      \begin{eqnarray}
 \sum_j M_{ij} \phi^n_j=E_n \phi^n_i 
\label{Mphi}
  \end{eqnarray}
 where
       \begin{eqnarray}
 M_{ij} = \left( 
 \begin{array}{cccc}
h_{i \uparrow, j \uparrow}  e^{ {i \over 2} ({\chi}_i- {\chi}_j)} & h_{i \uparrow, j \downarrow} e^{ {i \over 2} ({\chi}_i- {\chi}_j)} & 0&  \Delta_{ij} \\
h_{i \downarrow, j \uparrow} e^{ {i \over 2} ({\chi}_i- {\chi}_j)} & h_{i \downarrow, j \downarrow} e^{ {i \over 2} ({\chi}_i- {\chi}_j)}&   \Delta_{ji}  & 0 \\
 0 & \Delta_{ij}^{\ast} & -h^{\ast}_{i \uparrow, j \uparrow} e^{ -{i \over 2} ({\chi}_i- {\chi}_j)} &   h^{\ast}_{i \uparrow, j \downarrow} e^{ -{i \over 2} ({\chi}_i- {\chi}_j)} \\
  \Delta_{ji}^{\ast} & 0 &  h^{\ast}_{i \downarrow, j \uparrow} e^{ -{i \over 2} ({\chi}_i - \chi_j)} & -h^{\ast}_{i \downarrow, j \downarrow} e^{ -{i \over 2} ({\chi}_i- {\chi}_j)} 
 \end{array}
 \right)
  \end{eqnarray}
  and 
         \begin{eqnarray}
 \phi^n_i = \left( 
 \begin{array}{cccc}
 u^n_{i \uparrow}\\
  u^n_{i \downarrow}\\
   v^n_{i \uparrow}\\
  v^n_{i \downarrow}
 \end{array}
 \right)
  \end{eqnarray}
  
  Note that solutions for $E_n$ and $-E_n$ are connected; the solution for $-E_n$ is given by
  \begin{eqnarray}
  \left( 
 \begin{array}{cccc}
 -(v^n_{i \uparrow})^{\ast}\\
  (v^n_{i \downarrow})^{\ast}\\
   -(u^n_{i \uparrow})^{\ast}\\
  (u^n_{i \downarrow})^{\ast}
 \end{array}
 \right)
  \end{eqnarray}
  Therefore, we only need to obtain half of the solutions to Eq.~(\ref{Mphi}).
  
  Self-consistent solutions are obtained by solving the above equations using the self-consistent fields, 
   \begin{eqnarray}
 \langle n_{i \uparrow} \rangle&=& \langle c^{\dagger}_{i \uparrow} c_{i \uparrow} \rangle =\sum_{n_1}'  \sum_{n_2}'
 \langle
 e^{ {i \over 2} \hat{\chi}_i}[ u^{n_1}_{i \uparrow}\gamma_{n_1}- (v^{n_1}_{i \uparrow})^{\ast}\gamma_{n_1}^{\dagger}]^{\dagger}
  [ u^{n_2}_{i \uparrow}\gamma_{n_2}- (v^{n_2}_{i \uparrow})^{\ast}\gamma_{n_2}^{\dagger}] e^{ -{i \over 2} \hat{\chi}_i}
 \rangle
  \nonumber
 \\
 &=& \sum_{n_1}'  \sum_{n_2}'
 \langle
 e^{ {i \over 2} {\chi}_i}[ (u^{n_1}_{i \uparrow})^{\ast} \gamma^{\dagger}_{n_1}- (v^{n_1}_{i \uparrow})\gamma_{n_1}]
  [ u^{n_2}_{i \uparrow}\gamma_{n_2}- (v^{n_2}_{i \uparrow})^{\ast}\gamma_{n_2}^{\dagger}] e^{ -{i \over 2} {\chi}_i}
 \rangle
  \nonumber
 \\
  &=& \sum_{n_1}'  \sum_{n_2}'
 [ (u^{n_1}_{i \uparrow})^{\ast} u^{n_2}_{i \uparrow} \langle\gamma^{\dagger}_{n_1}\gamma_{n_2}\rangle
 -(u^{n_1}_{i \uparrow})^{\ast} (v^{n_2}_{i \uparrow})^{\ast}\langle\gamma^{\dagger}_{n_1}\gamma_{n_2}^{\dagger}\rangle + (v^{n_1}_{i \uparrow})u^{n_2}_{i \uparrow} \langle \gamma_{n_1} \gamma_{n_2}\rangle + (v^{n_1}_{i \uparrow}) (v^{n_2}_{i \uparrow})^{\ast} \langle \gamma_{n_1}\gamma_{n_2}^{\dagger}\rangle]
  \nonumber
 \\
 &=& \sum_n' [| u^{n}_{i \uparrow}|^2 f(E_n)+|v^{n}_{i \uparrow}|^2 f(-E_n)]= \sum_n | u^{n}_{i \uparrow}|^2 f(E_n)= \sum_n | v^{n}_{i \uparrow}|^2 f(-E_n)
   \nonumber
 \\
  \langle n_{i \downarrow} \rangle &=& \sum_n' [| u^{n}_{i \downarrow}|^2 f(E_n)+|v^{n}_{i \downarrow}|^2 f(-E_n)]
  = \sum_n | u^{n}_{i \downarrow}|^2 f(E_n)= \sum_n | v^{n}_{i \downarrow}|^2 f(-E_n)
  \label{eq144}
  \end{eqnarray}
    \begin{eqnarray}
  \langle { S}^x_{i} \rangle &=&-{ 1 \over 2} \sum_{n}' [ v^{n}_{i \uparrow} (v^{n}_{i \downarrow})^{\ast} + {\rm c.c} ]f(-E_n)
  +{ 1 \over 2} \sum_{n}' [ u^{n}_{i \uparrow} (u^{n}_{i \downarrow})^{\ast} + {\rm c.c} ]f(E_n)
   \nonumber
 \\
  \langle { S}^y_{i} \rangle &=&{i\over 2} \sum_{n}' [ v^{n}_{i \uparrow} (v^{n}_{i \downarrow})^{\ast} - {\rm c.c} ]f(-E_n)
 - {i\over 2} \sum_{n}' [ u^{n}_{i \uparrow} (u^{n}_{i \downarrow})^{\ast} - {\rm c.c} ]f(E_n)
   \nonumber
 \\
 \langle { S}^z_{i} \rangle &=&{ 1 \over 2} \sum_{n}' [ v^{n}_{i \uparrow} (v^{n}_{i \uparrow})^{\ast} - v^{n}_{i \downarrow} (v^{n}_{i \downarrow})^{\ast} ]f(-E_n)
 +{ 1 \over 2} \sum_{n}' [ u^{n}_{i \uparrow} (u^{n}_{i \uparrow})^{\ast} - u^{n}_{i \downarrow} (u^{n}_{i \downarrow})^{\ast} ]f(E_n)
   \label{eq147}
  \end{eqnarray}
  and
        \begin{eqnarray}
 \Delta_{i j}= {{2t_1^2} \over U}  \sum'_n [ u^{n}_{i \uparrow} (v^{n}_{j \downarrow})^{\ast} +u^{n}_{j \downarrow} (v^{n}_{i \uparrow})^{\ast}+  u^{n}_{i \downarrow} (v^{n}_{j \uparrow})^{\ast}+ u^{n}_{j \uparrow} (v^{n}_{i \downarrow})^{\ast}]
 \tanh {{E_{n}} \over {2 k_B T}}
    \label{eq148}
  \end{eqnarray}
where following relations are used
          \begin{eqnarray}
    \langle \gamma^{\dagger}_n \gamma_n  \rangle =f(E_n), \quad    \langle \gamma_n  \gamma^{\dagger}_n  \rangle =1-f(E_n)=f(-E_n),
    \end{eqnarray}

 We solve the system of equations in Eq.~(\ref{Mphi}) by the Car-Parrinello method \cite{Car}, taking $\phi^n_i$ as ``time-dependent'' variable $\phi^n(t)$ at the $i$th site that follows the Newtonian dynamics
 \begin{eqnarray}
 m_{\rm CP} \ddot{\phi}^n_i(t)= -\sum_j M_{ij} \phi^n_j(t)+E_n \phi^n_i(t) -\eta_{\rm CP}\dot{\phi}^n_i(t)
 \label{CP}
 \end{eqnarray}
 where  $m_{\rm CP}$ and $\eta_{\rm CP}$ are mass and friction coefficient for the variable, respectively; here, the wave functions $\{ \phi^n(t) \}$ are orthonormalized during the calculation.

\subsection{Spin-vortices and multi-valued wave functions}

The self-consistent wave functions obtained in the previous section, actually, become multi-valued with respect to electron coordinates due to the presence of spin-vortices. The spin-texture obtained for a model system composed of the two-layers shown in Fig.~\ref{Lattice2} is depicted in Fig.~\ref{Spins}.

If $ u^{n}_{i \sigma}$ and $v^{n}_{i \sigma}$ are obtained just by minimizing energy, 
the resulting state is a currentless state that satisfies the so-called ``Bloch's theorem'' \cite{Bohm-Bloch}.
This corresponds to the solution with constant $\chi$.

Let us examine this currentless state, more closely.
Replacing the operator $\hat{\chi}_i$ in Eq.~(\ref{eq137}) by the scalar value of the underlying Berry connection, we obtain
 \begin{eqnarray}
    \left(
    \begin{array}{c}
 c_{ i \uparrow} \\
  c_{ i \downarrow} 
 \end{array}
 \right)
 =e^{ -{i \over 2} {\chi}_i}
\sum_{n}' \left(
 \begin{array}{cc}
  u^{n}_{i \uparrow}   & - (v^{n}_{i \uparrow})^{\ast}\\
    u^{n}_{i \downarrow} &( v^{n}_{i \downarrow})^{\ast}
    \end{array}
    \right)
    \left(
    \begin{array}{c}
\gamma_{n} \\
  \gamma^{\dagger}_{n} 
 \end{array}
 \right)
 \label{multi-value}
  \end{eqnarray}
and the currentless result corresponds to the solution with constant $\chi_j$. 
Since the single-valued solution corresponds to multi-valued $\chi_j$, the currentless $u^{n}_{i \sigma}$ and $v^{n}_{i \sigma}$ are actually multi-valued. 

When the bulk Hamiltonian yields spin-vortices lying in the $xy$ plane (in the CuO$_2$ plane),  the on-site $U$ term in $H^{\rm HF}_{\rm EHFS}$ in Eq.~(\ref{SV2:H^HF_EHFS-2})
becomes
\begin{eqnarray}
U \sum_j [\frac{1}{2}(c^\dagger_{j\uparrow} c_{j\uparrow}
     +c^\dagger_{j\downarrow} c_{j\downarrow} )
     -\frac{2}{3}\left|\langle {\bf S}_j \rangle \right|(e^{-i\xi_j} c^\dagger_{j\uparrow} c_{j\downarrow}
       +e^{i\xi_j} c^\dagger_{j\downarrow} c_{j\uparrow})
     - \frac{2}{3} \langle{\bf S}_j\rangle^2 ]
\end{eqnarray}

We can abosrbe the spin-twisting factor $e^{\pm i\xi_j}$ using $\tilde{c}^{\dagger}_{j\uparrow}, \tilde{c}_{j\downarrow}$ as
\begin{eqnarray}
 c^{\dagger}_{j \uparrow}  = e^{{i \over 2} \xi_j}\tilde{c}^{\dagger}_{j\uparrow}, \quad
c_{j \downarrow}  = e^{{i \over 2} \xi_j}\tilde{c}_{j\downarrow}
\label{eq152}
\end{eqnarray}
Then, Eq.~(\ref{multi-value}) becomes
\begin{eqnarray}
    \left(
    \begin{array}{c}
 \tilde{c}_{ i \uparrow} \\
  \tilde{c}_{ i \downarrow} 
 \end{array}
 \right)
 =
\sum_{n}' \left(
 \begin{array}{cc}
  \tilde{u}^{n}_{i \uparrow}   & - (\tilde{v}^{n}_{i \uparrow})^{\ast}\\
    \tilde{u}^{n}_{i \downarrow} &( \tilde{v}^{n}_{i \downarrow})^{\ast}
    \end{array}
    \right)
    \left(
    \begin{array}{c}
\gamma_{n} \\
  \gamma^{\dagger}_{n} 
 \end{array}
 \right)
  \end{eqnarray}
  This is essentially equivalent to the solution with constant $\chi$ in Eq.~(\ref{multi-value}).
  This means that $\tilde{u}^{n}_{i \sigma}$ and $\tilde{v}^{n}_{i \sigma}$ are obtained from
  the energy minimization only calculation if ${c}^{\dagger}_{j\uparrow}, {c}_{j\downarrow}$ are replaced by $\tilde{c}^{\dagger}_{j\uparrow}, \tilde{c}_{j\downarrow}$. We do this in the following.

We obtain
$\tilde{u}^{n}_{i \sigma}$ and $\tilde{v}^{n}_{i \sigma}$ from the following 
Hamiltonian, in which  ${c}^{\dagger}_{j\sigma}$ and ${c}_{j\sigma}$ are replaced by $\tilde{c}^{\dagger}_{j\sigma}, \tilde{c}_{j\sigma}$, 
 \begin{eqnarray}
 \sum_j \tilde{M}_{ij} \tilde{\phi}^n_j=\tilde{E}_n \tilde{\phi}^n_i 
 \label{eq154}
  \end{eqnarray}
 where $\tilde{M}_{ij}$ and $\tilde{\phi}^n_i$ are given by
\begin{eqnarray}
 \tilde{M}_{ij} = \left( 
 \begin{array}{cccc}
h_{i \uparrow, j \uparrow}  e^{ {i \over 2} ({\xi}_i- {\xi}_j)} & h_{i \uparrow, j \downarrow} e^{ -{i \over 2} ({\xi}_i + {\xi}_j)} & 0&  \Delta_{ij}e^{ {i \over 2} ({\xi}_i- {\xi}_j)}  \\
h_{i \downarrow, j \uparrow} e^{-{i \over 2} ({\xi}_i+{\xi}_j)} & h_{i \downarrow, j \downarrow} e^{ -{i \over 2} ({\xi}_i- {\xi}_j)}&   \Delta_{ji} e^{- {i \over 2} ({\xi}_i- {\xi}_j)}  & 0 \\
 0 & \Delta_{ij}^{\ast}e^{ -{i \over 2} ({\xi}_i- {\xi}_j)}  & -h^{\ast}_{i \uparrow, j \uparrow} e^{- {i \over 2} ({\xi}_i- {\xi}_j)} &   h^{\ast}_{i \uparrow, j \downarrow} e^{ -{i \over 2} ({\xi}_i+{\xi}_j)} \\
  \Delta_{ji}^{\ast}e^{ {i \over 2} ({\xi}_i- {\xi}_j)}  & 0 & h^{\ast}_{i \downarrow, j \uparrow} e^{{i \over 2} ({\xi}_i+{\xi}_j)}& -h^{\ast}_{i \downarrow, j \downarrow} e^{ {i \over 2} ({\xi}_i- {\xi}_j)} 
 \end{array}
 \right)
  \end{eqnarray}
  and
   \begin{eqnarray}
 \tilde{\phi}^n_i = \left( 
 \begin{array}{cccc}
 \tilde{u}^n_{i \uparrow}\\
  \tilde{u}^n_{i \downarrow}\\
   \tilde{v}^n_{i \uparrow}\\
  \tilde{v}^n_{i \downarrow}
 \end{array}
 \right)
  \end{eqnarray}
respectively. Note that $\tilde{u}^{n}_{i \sigma}$, $\tilde{v}^{n}_{i \sigma}$ are single-valued since the multi-valued part $e^{ \pm {i \over 2} \xi}$ are removed.
 
  \subsection{Evaluation of Berry connection}
  
  Using $\tilde{u}^{n}_{i \sigma}$, $\tilde{v}^{n}_{i \sigma}$, Eq.~(\ref{multi-value}) is rewritten as 
 \begin{eqnarray}
    \left(
    \begin{array}{c}
 c_{ i \uparrow} \\
  c_{ i \downarrow} 
 \end{array}
 \right)
 =
\sum_{n}' \left(
 \begin{array}{cc}
  e^{ -{i \over 2} (\chi_i+\xi_i)} \tilde{u}^{n}_{i \uparrow}   & -  e^{ -{i \over 2} (\chi_i+\xi_i)}(\tilde{v}^{n}_{i \uparrow})^{\ast}\\
     e^{ -{i \over 2} (\chi_i-\xi_i)}\tilde{u}^{n}_{i \downarrow} & e^{ -{i \over 2} (\chi_i-\xi_i)}( \tilde{v}^{n}_{i \downarrow})^{\ast}
    \end{array}
    \right)
    \left(
    \begin{array}{c}
\gamma_{n} \\
  \gamma^{\dagger}_{n} 
 \end{array}
 \right)
 \label{multi-value2}
  \end{eqnarray}

Now, we need to obtain the multi-valued $\chi$.
The angular variable $\chi$ is obtained from conditions,
 \begin{eqnarray}
w_{C_{\ell}}[\xi]+ w_{C_{\ell}}[\chi] = \mbox{even number}  \mbox{  for any loop $C_{\ell}$}
\label{eqSingle}
\end{eqnarray}
If the above conditions are satisfied, $ e^{ -{i \over 2} (\chi_i\pm\xi_i)}$ appeared in Eq.~(\ref{multi-value2}) become single-valued.

Due to the presence of spin-vortices $\xi$ is multi-valued, and contains jump-of-value points of an integral multiple of $2\pi$. 
To establish the condition in Eq.~(\ref{eqSingle}), $\xi$ and $\chi$ must have the same jump-of-value points.

\begin{figure}
\begin{center}
\includegraphics[width=12.0cm]{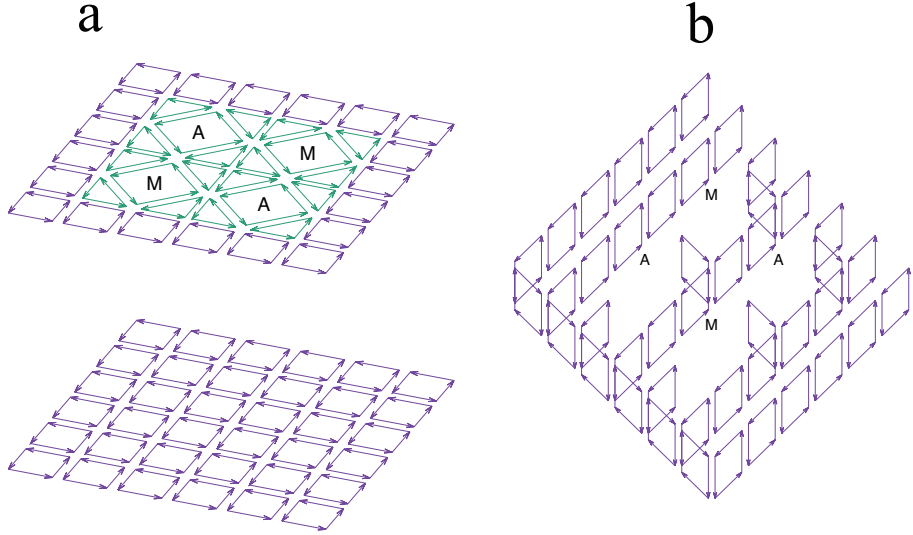}
\end{center}
\caption{Loops used to impose boundary conditions in Eq.~(\ref{eqSingle}). {\bf a}: loops in the bulk and surface layers. 
{\bf a}: loops in the bulk-surface interface region. Some of the loops in the system are removed since they are not independent; as a consequence,
all the cells are accessible from outside without breaking any walls framed by loops.
 }
\label{LoopsBC}
\end{figure}

To have the same jump-of-value points for $\xi$ and $\chi$, we do the following: 
first, we rebuild $\xi$ from differences of their values between bonds; $\xi_j$'s are obtained from the expectation values of the spin components $S^x_j$ and $S^y_j$; we choose a particular branch for $\xi_j$. 
The system we consider below has the antiferromagnetic background; this gives rise to the background value of $\xi$, $\xi^0_{j}=\pi (j_x+j_y+j_z)$ in the units the lattice constant, where $(j_x, j_y, j_z)$ is the $xyz$ coordinates of the $j$th site.

In order to separate the antiferromagnetic contribution from $\xi$, we introduce angular variable $\eta$, 
 \begin{eqnarray}
 \eta_j=\xi_j-\pi (j_x+j_y+j_z)
 \label{eta}
 \end{eqnarray}
  where $\eta_j$ is $\eta$ at the $j$th site.
 We take the branch of $\eta_j$ that satisfies the following condition, 
\begin{eqnarray}
-\pi \le \eta_{j} -\eta_k < \pi
\end{eqnarray}
where $k$ is a nearest neighbor site of $j$.

From ($\eta_j-\eta_k$)'s, we construct  ($\xi_j-\xi_k$)'s. After ($\xi_j-\xi_k$)'s are obtained, we rebuild $\xi$ from them. 
The process is as follows: first, we pick a value for the initial $\xi_1$. 
After fixing the value of $\xi_1$, we calculate $\xi_2$ by $\xi_{2} = \xi_1 + (\xi_{2} -\xi_1)$, where the site $2$ is connected to the site $1$ by a nearest neighbor bond.
 The step where value $\xi_{j}$ is derived from the already evaluated value of $\xi_k$ is given by
\begin{eqnarray}
\xi_{j} = \xi_k + (\xi_{j} -\xi_k)
\end{eqnarray}
where the sites ${j}$ and $k$ are connected by a bond in the path for the rebuilding of $\xi$. 
This process is continued until values at all accessible sites are evaluated once and only once. 

By this rebuilding process, a single path is constructed from the site $1$ to other sites $k\neq 1$, which is achieved by making the region singly-connected by removing some bonds (see Fig.~\ref{Lattice2}).
We denote the path from the site $1$ to other sites $k\neq 1$ by $C_{1 \rightarrow k}$.
Then, the value $\xi_k$ is given as
\begin{eqnarray}
\xi_k \approx \xi_1+ \int_{C_{1 \rightarrow k}} \nabla \xi \cdot d{\bf r}
\label{rebuilxi}
\end{eqnarray}

Let us summarize the calculation so far.

\begin{enumerate}

\item Solve Eq.~(\ref{Mphi}) self-consistently using the Car-Parrinello method in Eq.~(\ref{CP}). Thereby, we obtain Eqs.~(\ref{eq144}), (\ref{eq147}), and (\ref{eq148}). 

\item From Eq.~(\ref{eq147}), $\xi_k$'s are obtained. Next, $\xi_k$'s are rebuilt using Eq.~(\ref{rebuilxi}).
Thereby, we have $\xi_k$'s for which we know the positions of the jump-of-values.  

\item From the rebuilt $\xi_k$'s, construct $\tilde{c}^{\dagger}_{j\sigma}, \tilde{c}_{j\sigma}$ in Eq.~(\ref{eq152}).
Using Eqs.~(\ref{eq144}), (\ref{eq147}), and (\ref{eq148}) obtained in Step 1),  solve Eq.~(\ref{eq154}). Thereby,
$\tilde{u}^{n}_{i \sigma}$, $\tilde{v}^{n}_{i \sigma}$ are obtained. 

\item $e^{\pm {i \over 2}\xi_i}$, $\tilde{u}^{n}_{i \sigma}$, $\tilde{v}^{n}_{i \sigma}$, ${u}^{n}_{i \sigma}$ and ${v}^{n}_{i \sigma}$ are obtained; $
{u}^{n}_{i \uparrow}= e^{- {i \over 2}\xi_i}\tilde{u}^{n}_{i \uparrow}$, ${u}^{n}_{i \downarrow}= e^{ {i \over 2}\xi_i}\tilde{u}^{n}_{i \downarrow}$,
${v}^{n}_{i \uparrow}= e^{{i \over 2}\xi_i}\tilde{v}^{n}_{i \uparrow}$, ${v}^{n}_{i \downarrow}= e^{ -{i \over 2}\xi_i}\tilde{v}^{n}_{i \downarrow}$.

\end{enumerate}

Next, we obtain $\chi_k$'s that satisfy Eq.~(\ref{eqSingle}). 
The requirement in Eq.~(\ref{eqSingle}) is given by
\begin{eqnarray}
w_{C_{\ell}}[\chi]={ 1 \over {2\pi}} \sum_{i=1}^{N_{\ell}}  \tau_{{C_{\ell}(i+1)} \leftarrow {C_{\ell}(i)}}
\label{eq164}
\end{eqnarray}
where $w_{C_{\ell}}[\chi]$ is the number supplied as a boundary condition. $C_{\ell}(i)$ is the $i$th site in the loop $C_{\ell}$, where
$C_{N_\ell+1}=C_{1}$ with $N_\ell$ is the number of sites in $C_{\ell}$.

In order to impose conditions in Eq.~(\ref{eq164}), $\tau_{ j \leftarrow i}$ is split into a multi-valued part $\tau_{ j \leftarrow i}^0$ and 
single-valued part $f_{ j \leftarrow i}$ as
\begin{eqnarray}
\tau_{ j \leftarrow i}=\tau_{ j \leftarrow i}^0 + f_{ j \leftarrow i}
\end{eqnarray}
where $\tau_{ j \leftarrow i}^0$ is given to satisfy
\begin{eqnarray}
w_{C_{\ell}}[\chi]={ 1 \over {2\pi}} \sum_{i=1}^{N_{\ell}}  \tau^0_{{C_{\ell}(i+1)} \leftarrow {C_{\ell}(i)}}
\end{eqnarray}
and $f_{ j \leftarrow i}$ is solved to satisfy
\begin{eqnarray}
0={ 1 \over {2\pi}} \sum_{i=1}^{N_{\ell}}  f_{{C_{\ell}(i+1)} \leftarrow {C_{\ell}(i)}}
\label{eq164}
\end{eqnarray}
The above is the equation corresponding to Eq.~(\ref{Feq2b}). They are imposed on independent loops in the system, as shown in Fig.~\ref{LoopsBC}.

For Eq.~(\ref{Feq1b}), we use the conservation of the local charge at sites instead. As will be explained later,
it gives rise to a sufficient number of equations to obtain unknowns.

The conservation of the local charge at the $k$th site is 
\begin{eqnarray}
0= \sum_{\ell \in b_k}J_{k \leftarrow \ell}
\label{Feq3}
\end{eqnarray}
where $b_k$ is a set of sites that are connected to the $k$th site.
 $J_{k \leftarrow \ell}$ is the current through the bond between sites $k$ and $\ell$ in the direction  $k \leftarrow \ell$, 
\begin{eqnarray}
J_{k \leftarrow \ell}&=&{{2e} \over \hbar}  {{\partial E} \over {\partial \tau_{ k \leftarrow \ell}}}={{2e} \over \hbar}  {{\partial } \over {\partial \tau_{ k \leftarrow \ell}}} \langle {\rm Gnd} | \sum_{i,j,\sigma, \sigma'} h_{i \sigma, j \sigma'} c_{i \sigma}^{\dagger} c_{j \sigma'} |{\rm Gnd} \rangle
\nonumber
\\
&=&{{2e} \over \hbar}  {{\partial } \over {\partial \tau_{ k \leftarrow \ell}}} \sum_{i,j,\sigma, \sigma'} h_{i \sigma, j \sigma'}
\sum_{n_1}'  \sum_{n_2}'
 \langle
 e^{ {i \over 2} \hat{\chi}_i}[ (u^{n_1}_{i \sigma})^{\ast}\gamma_{n_1}^{\dagger}-\sigma (v^{n_1}_{i \sigma})\gamma_{n_1}]
  [ u^{n_2}_{j \sigma'}\gamma_{n_2}- \sigma'(v^{n_2}_{j \sigma'})^{\ast}\gamma_{n_2}^{\dagger}] e^{ -{i \over 2} \hat{\chi}_j}
 \rangle
\nonumber
\\
&=&{{2e} \over \hbar}   \sum_{i,j,\sigma, \sigma'} h_{i \sigma, j \sigma'}
\sum_{n}' 
 {{\partial e^{ {i \over 2} \tau_{i \leftarrow j}} } \over {\partial \tau_{ k \leftarrow \ell}}}\left[ (u^{n}_{i \sigma})^{\ast}(u^{n}_{j \sigma'})f(E_n)
 +\sigma \sigma'  (v^{n}_{i \sigma})(v^{n}_{j \sigma'})^{\ast}f(-E_n) \right]
\nonumber
\\
&=&{{2e} \over \hbar}\sum_{n, i,j,\sigma}' {{\partial e^{ {i \over 2} \tau_{i \leftarrow j}} } \over {\partial \tau_{ k \leftarrow \ell}}}
\Big\{ h_{i \sigma, j \sigma} [(u^n_{i \sigma})^{\ast}u^n_{j \sigma}f(E_n)+ (v^n_{j \sigma})^{\ast}v^n_{i \sigma}f(-E_n)]
\nonumber
\\
&+&
h_{i \sigma, j -\sigma} [(u^n_{i \sigma})^{\ast}u^n_{j -\sigma}f(E_n)- (v^n_{j -\sigma})^{\ast}v^n_{i \sigma}f(-E_n)]
 \Big\}
\end{eqnarray}

We employ an iterative improvement of the approximate solutions by using the linearized version of Eq.~(\ref{Feq3}) given by
\begin{eqnarray}
0 \approx  {{2e} \over \hbar} \sum_i  {{\partial E (\{ \tau_{ j \leftarrow i}^0 \}) } \over {\partial \tau_{ j \leftarrow i}}}+{{2e} \over \hbar} 
\sum_i  {{\partial^2 E (\{ \tau_{ j \leftarrow i}^0 \}) } \over {\partial (\tau_{ j \leftarrow i})^2}}f_{ j \leftarrow i}
\label{Feq4}
\end{eqnarray}

A system of equations for $f_{ j \leftarrow i}$'s composed of Eqs.~(\ref{eq164}) and (\ref{Feq4}) are solve for given $\tau_{ j \leftarrow i}^0$'s by  updating them iteratively 
\begin{eqnarray}
\tau_{ j \leftarrow i}^{ 0 \ 
New}=\tau_{ j \leftarrow i}^{ 0 \ Old}+f_{ j \leftarrow i}
\end{eqnarray}
where $\tau_{ j \leftarrow i}^{ 0 \ Old}$ is $\tau_{ j \leftarrow i}^{ 0}$ value that is used to obtain the current value of $f_{ j \leftarrow i}$; $\tau_{ j \leftarrow i}^{ 0 \ New}$ will be used to obtain the next $f_{ j \leftarrow i}$ value. 

The numerical convergence is checked by the condition
\begin{eqnarray}
\left| {{2e} \over \hbar} \sum_i  {{\partial E (\{ \tau_{ j \leftarrow i}^0 \}) } \over {\partial \tau_{ j \leftarrow i}}} \right| < \epsilon
\end{eqnarray}
where $\epsilon$ is a small number.

For the initial $\tau_{ j \leftarrow i}^{ 0 }$, we adopt the following,
\begin{eqnarray}
\tau_{ j \leftarrow i}^{ 0 \ init}=\sum_h  w_h \tan^{-1} {{ j_y- h_y} \over {j_x -h_x}}-\sum_h w_h \tan^{-1} {{ i_y- h_y} \over {i_x -h_x}}
\end{eqnarray}
where $(j_x,j_y,j_z)$ and $(i_x,i_y,i_z)$ are  coordinates of the sites $j$ and $i$, respectively, $h=(h_x,h_y,h_z)$ is the coordinate of the hole occupied site, and $w_h$ is the winding number of $\chi$ around the hole at $h$.

The number of $\tau_{ j \leftarrow i}$ to be evaluated is equal to the number of the bonds.
The number of equations in Eq.~(\ref{Feq2b}) is equal to the number of the plaques.
The number of equations from Eq.~(\ref{Feq3}) for the conservation of charge is equal to the number of sites$-1$, due to the fact that the total charge is fixed in the calculation.

The equality of the number of unknowns and the number of equations gives
\begin{eqnarray}
[\mbox{\# bonds}]=[\mbox{\# plaques}]+[\mbox{\# sites}-1]
\label{Euler1}
\end{eqnarray}

 It is interesting to note that this agrees with the Euler's theorem for the two-dimensional lattice 
\begin{eqnarray}
[\mbox{\# edges}]=[\mbox{\# faces}]+[\mbox{\# vertices}-1]
\label{Euler2}
\end{eqnarray}

\begin{figure}
\begin{center}
\includegraphics[width=10.0cm]{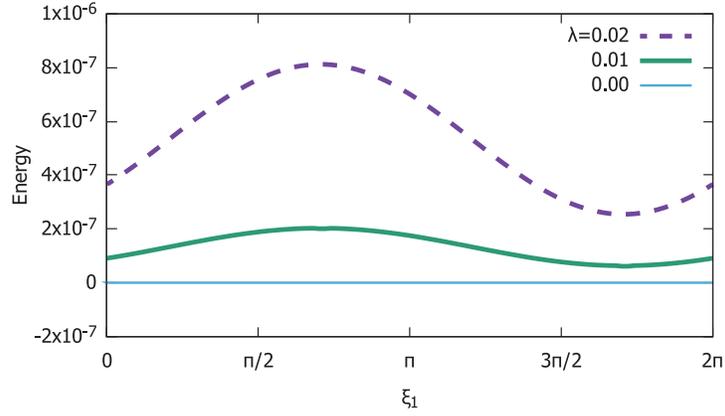}
\end{center}
\caption{Dependence of the total energy on the value of $\xi_1$. The zero of energy is taken to be the value for $\lambda=0$ calculation.
The units of the energy is $t_1$.}
\label{xi1}
\end{figure}

\begin{figure}
\begin{center}
\includegraphics[width=16.0cm]{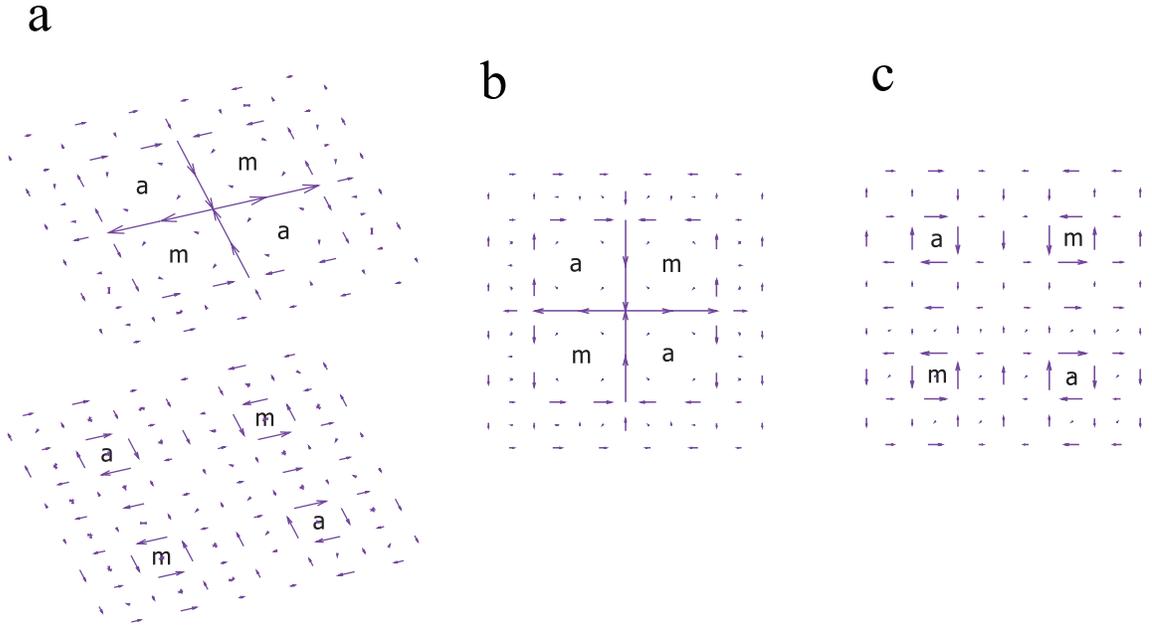}
\end{center}
\caption{Loop current obtained by the calculation. {\bf a}: Loop current in the system. Each ``m'' indicates a center of a loop current with winding number $+1$,
and ``a'' a center of a loop-current with winding number $-1$. {\bf b}: Loop current in the bulk layer viewed from above. {\bf c}: Loop current in the surface layer viewed from above.}
\label{current}
\end{figure}

 When the spin-orbit interaction exists the total energy depends on $\xi_1$
 as shown in Fig.~\ref{xi1}. In Fig.~\ref{current}, obtained current using $\xi_1=0$ is depicted. 
 
\subsection{Diamagnetic current produced by an application of a magnetic field}

\begin{figure}
\begin{center}
\includegraphics[width=16.0cm]{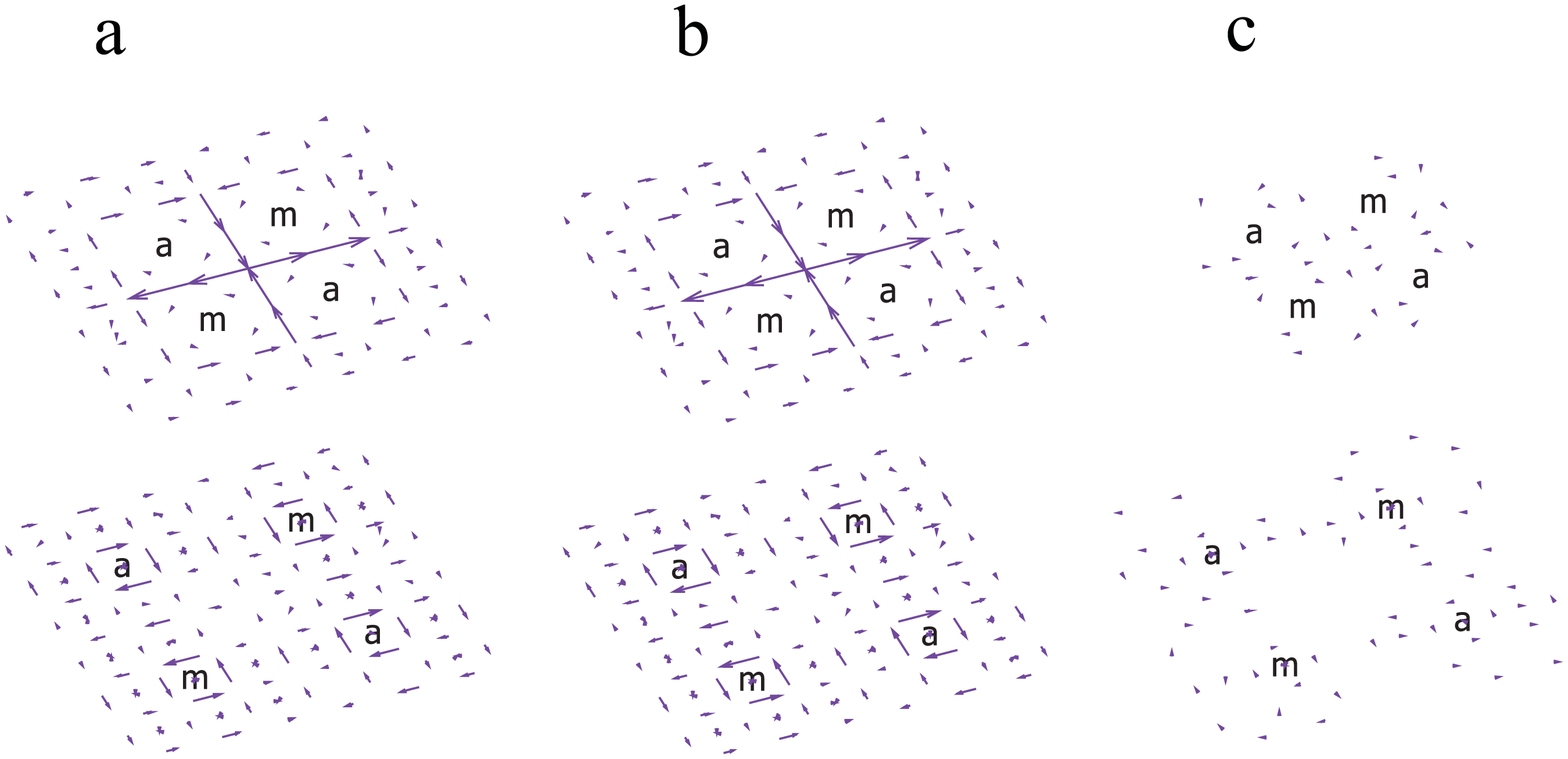}
\end{center}
\caption{Supercurrent under the application of an external magnetic field. $B=0.01$ in the units of $t_1=1, \hbar=1, e=1$, and lattice constant in the plane $=1$. {\bf a}: The result from the calculation.  {\bf b}: The linear approximation
given in Eq.~(\ref{eqLondonAp}). {\bf c}: The difference between the calculated result and its linear approximation.}
\label{jAem}
\end{figure}

 One of the hallmarks of superconducting states is the diamagnetic response to an external magnetic field.
 Let us consider the situation where a magnetic field ${\bf B}^{\rm em}=\nabla \times {\bf A}^{\rm em}$ is applied.
Then, the modified energy functional in Eq.~(\ref{energyfAem}) is used. 

This leads to replace $\tau_{ j \leftarrow i}$ in $E$ by
\begin{eqnarray}
u_{ j \leftarrow i}=\tau_{ j \leftarrow i}-{{2e} \over {\hbar c}} \int^j_i{\bf A}^{\rm em}\cdot d{\bf r}
\end{eqnarray}
where integration is performed along the bond $ j \leftarrow i$.

The calculation can be done similarly to the case for no magnetic field, starting from the initial value
\begin{eqnarray}
u_{ j \leftarrow i}^{0 \ init}=\tau_{ j \leftarrow i}^{ 0  \ init}-{{2e} \over {\hbar c}} \int_{j \leftarrow i}{\bf A}^{\rm em}\cdot d{\bf r}
\end{eqnarray}

Note that during the evaluation process of $\nabla \chi$, the ambiguity in the gauge of ${\bf A}^{\rm em}$ is compensated, thus, the effective vector potential 
\begin{eqnarray}
{\bf A}^{\rm eff}= {\bf A}^{\rm em}-{{\hbar c}  \over {2e}} \nabla \chi
\end{eqnarray}
is invariant with respect to the choice of the gauge in ${\bf A}^{\rm em}$. 

Now we apply a uniform magnetic field perpendicular to the lattice.
In the actual numerical calculations we have adopted
\begin{eqnarray}
{\bf A}^{\rm em}=
\left(
\begin{array}{c}
-By \\
0 \\
0
\end{array}
\right)
\end{eqnarray}
but we checked that the calculated current distribution is identical even other gauge is employed.

In Fig.~\ref{jAem}{\bf a}, current distributions with applying the magnetic field are depicted.

Numerical calculations indicate the following relation holds
\begin{eqnarray}
J_{j \leftarrow i}\approx -{{4 e^2} \over \hbar^2} { {\partial^2 E[\{0\}]} \over {\partial (u_{j \leftarrow i})^2} } \int_i^j {\bf A}^{\rm eff} \cdot d {\bf r}
\label{eqLondonAp}
\end{eqnarray}
where $\{0\}$ means all $u_{j \leftarrow i}$'s are zero \cite{Manabe2019}. 
The result using the above approximation is shown in Fig.~\ref{jAem}{\bf b}, and the difference between the 
exact result and approximate one is shown in Fig.~\ref{jAem}{\bf c}.
The approximate one is almost identical to the exact one, except the current just around the small polarons; it originates directly from the Rashba interaction.

\begin{figure}
\begin{center}
\includegraphics[width=16.0cm]{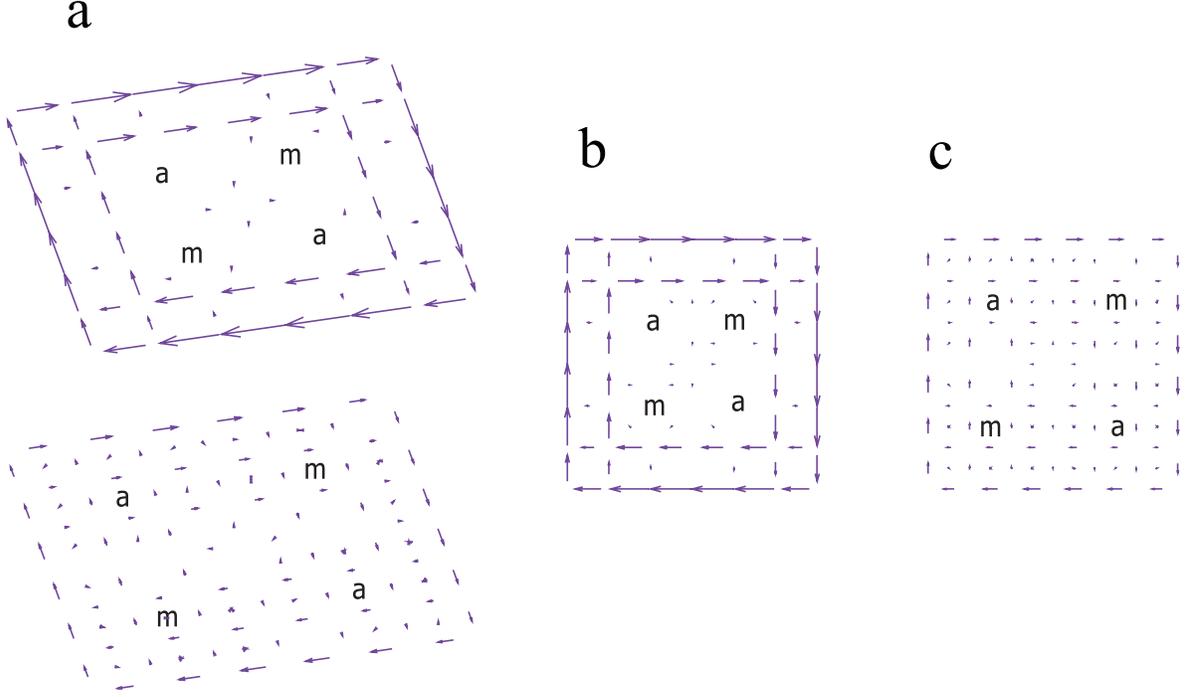}
\end{center}
\caption{Supercurrent indued by the application of an external magnetic field.  {\bf a}: The induced current in the two layers.  The scale for the current arrow is enlarged from Fig.~\ref{jAem}. {\bf b}: The induced current in the bulk layer. {\bf c}: 
the induced current in the surface layer.}
\label{jAem2}
\end{figure}

The induced current is diamagnetic as seen in Fig.~\ref{jAem2}. This will become a screening current of the magnetic field in a
large system size.

\subsection{Zero voltage current production by an external current feeding}

Another hallmark of superconducting states is the zero voltage current flow through the sample.

Let us consider the external current feeding.
When external currents are fed, the conservation of the local charge at site $k$ in Eq.~(\ref{Feq3}) is modified as
\begin{eqnarray}
0= \sum_{\ell \in b_k}J_{k \leftarrow \ell}+J^{\rm EX}_{k}
\label{mFeq3}
\end{eqnarray}
where $J^{\rm EX}_{k}$ is the external current fed at $k$.

\begin{figure}
\begin{center}
\includegraphics[width=12.0cm]{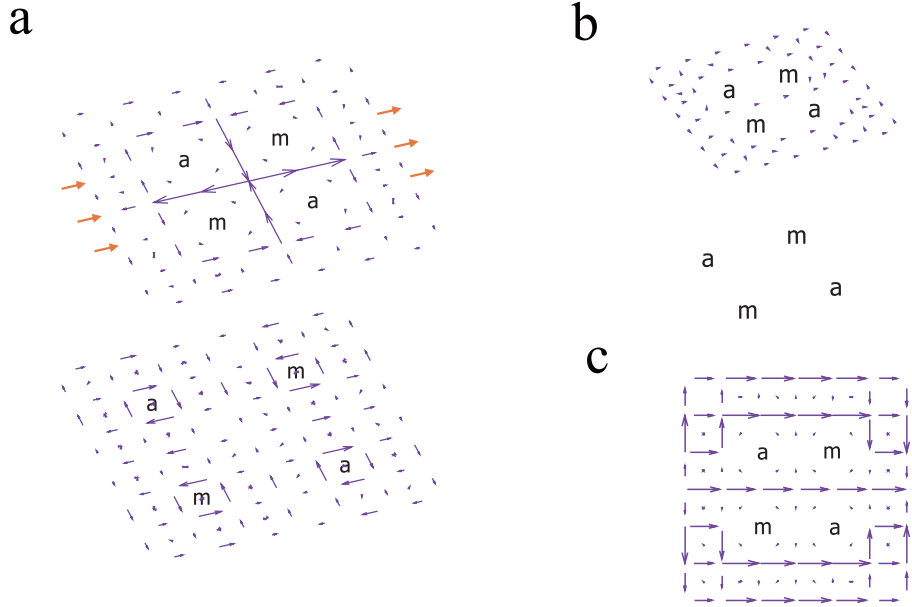}
\end{center}
\caption{Supercurrent obtained by feeding external current. {\bf a}: Supercurrent with external current. Two sets of three arrows indicate
sites where external current enters and exits.  Each arrow indicates $J^{\rm EX}$. The units of the current is $t_1=1, \hbar =1$, and $e=1$. {\bf b}: The difference of currents with and without external current.
{\bf c}: The difference of currents with and without external current in the bulk layer viewed from above. Arrows are magnified from those in {\bf b}.}
\label{jex}
\end{figure}

\begin{figure}
\begin{center}
\includegraphics[width=10.0cm]{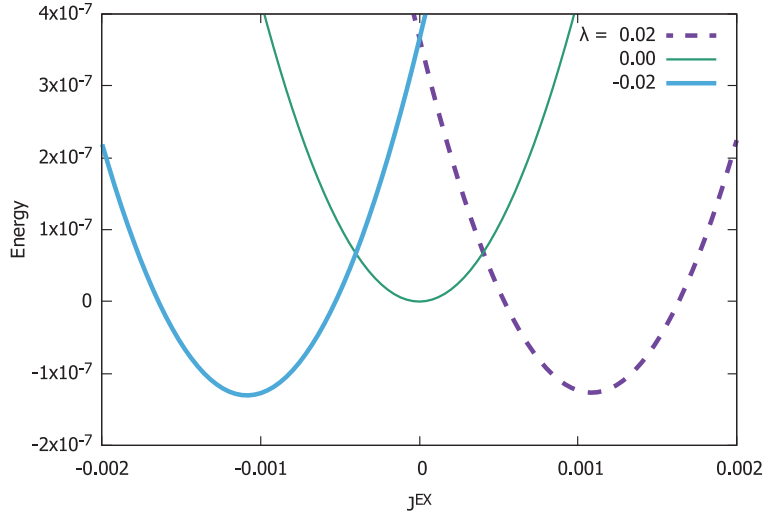}
\end{center}
\caption{Total energy vs $J^{\rm EX}$. The units of $J^{\rm EX}$ is $t_1=1, \hbar =1$, and $e=1$. Three different  Rashba parameters $\lambda$ are employed. The zero of the energy is that of $\lambda=0$ and $J^{\rm EX}=0$. }
\label{lm}
\end{figure}

The results are shown in Fig.~\ref{lm}. The energy minimum with nonzero $J^{\rm EX}$ occurs for $\lambda \neq0$ cases. This situation is similar to the one envisaged by Bloch for superconducting states \cite{Bloch1966}.  When the system is at the energy minimum with nonzero $J^{\rm EX}$, 
zero voltage current production is realized. In other words, persistent current flows through the sample.

The energy minimum occurs at zero $J^{\rm EX}$ for the $\lambda =0$ case. This indicates the if the Rashba interaction is absent,
 the zero voltage current production does not occur  in the present model.

\section{Concluding remarks}
\label{sec6}

We have presented a fundamentally revised theory for superconductivity, and the method to calculate properties of superconducting states based on it.

Since the early work of London, it has been repeatedly argued that theory of superconductivity needs to have a $U(1)$ gauge field different from the electromagnetic one to have the gauge invariant supercurrent. London called it ``superpotential'', and the
current standard theory calls it ``Nambu-Goldston mode''. In the new theory, the Berry connection ${\bf A}^{\rm fic}$ provides the necessary gauge field.

 We would like to emphasize that the vector potential provided by the Berry connection ${\bf A}^{\rm fic}$ plays a crucial role in realizing reversible superconducting-normal metal phase transition in a magnetic field \cite{koizumi2020b,koizumi2020c,koizumi2021,koizumi2021b}.
 This phenomenon cannot be explained by the standard theory \cite{Hirsch2020,koizumi2020b}.

Besides the reversible superconducting-normal metal phase transition in a magnetic field, the new theory solves all four problems described in Introduction. It also provides a calculation framework for the two of the hallmarks of superconducting states, the diamagnetic response current to an external magnetic field, and the zero voltage current produced by an external current feeding, microscopically. This theory may be able to elucidate the
cuprate superconductivity, where nanoscale inhomogeneity of electronic states is important.

Lastly, we would like to note that a similar formalism using the Berry connection is possible for superfluidity of bosons \cite{koizumi2019}.

\appendix

\section{Berry connection from many-body wave functions and the modification of Maxwell's equations}
\label{Appendix1}

Berry's derivation for the Berry phase uses the adiabatic assumption \cite{Berry}. The adiabatic assumption here does not necessary means a slow change. We will explain this point below.

By following Berry, let us consider the state vector where given by
\begin{eqnarray}
 |n({\bf R}(t)) \rangle
\end{eqnarray}
where ${\bf R}$ is the adiabatic parameter. This indicates the parametric time-dependence occrs through time-dependent adiabatic parameter ${\bf R}(t)$.

Then, the time-dependent wave function becomes
\begin{eqnarray}
 |\Psi(t) \rangle&=& e^{i \int_{{\bf R}(0)}^{{\bf R}(t)} \langle n({\bf R}(t))|\nabla_{\bf R}| n({\bf R}(t)) \rangle \cdot d{\bf R}}|n({\bf R}(t)) \rangle e^{-{i \over \hbar} \int^t_0 dt' E_n(t')}
 \nonumber
 \\
 &=& |n({\bf R}(t)) \rangle e^{i \gamma_n(t)} e^{-{i \over \hbar} \int^t_0 dt' E_n(t')}
 \label{BerryWf}
\end{eqnarray}

The phase
\begin{eqnarray}
\gamma_n(t)=i\int_{{\bf R}(0)}^{{\bf R}(t)} d{\bf R}\cdot {\bf A}_n({\bf R})
\end{eqnarray}
is the Berry phase, and 
\begin{eqnarray}
{\bf A}_n({\bf R})=-i\langle n({\bf R}) |\nabla_{\bf R} |n({\bf R}) \rangle 
\end{eqnarray}
is the Berry connection.

The Berry connection gives rise to a vector potential on the dynamics of ${\bf R}$.
Let us see this below, taking the Born-Oppenheimer approximation as an example.

In the Born-Oppenheimer approximation, ${\bf R}$ is the nuclear coordinates. The momentum conjugate to ${\bf R}$ is $-i\hbar \nabla_{\bf R}$.
The electron coordinates is denoted by ${\bf r}$.  
 The Born-Oppenheimer total wave function we consider is 
 \begin{eqnarray}
 f({\bf R}) \langle {\bf r}|n({\bf R}) \rangle
 \end{eqnarray}
where $f({\bf R})$ is the nuclear wave function and $\langle {\bf r}|n({\bf R}) \rangle$ is the electronic wave function with the nuclear coordinate as the adiabatic parameter.

Application of the momentum operator conjugate to ${\bf R}$ on the total wave function yields
 \begin{eqnarray}
 -i\hbar \nabla_{\bf R} \left[  f({\bf R})  \langle {\bf r}|n({\bf R}) \rangle \right]= \langle {\bf r}|n({\bf R}) \rangle[-i\hbar \nabla_{\bf R}f({\bf R}) ]
 - f({\bf R}) [i\hbar \nabla_{\bf R} \langle {\bf r}|n({\bf R}) \rangle]
 \label{eqgauge}
 \end{eqnarray}
 
The integration over ${\bf r}$ after applying $\langle n({\bf R})|{\bf r} \rangle$ from the left yields
\begin{eqnarray}
[-i\hbar \nabla_{\bf R} +\hbar {\bf A}_n({\bf R})]f({\bf R})
\end{eqnarray}
This indicates that the dynamics of ${\bf R}$ is that in the gauge field with the vector potential $\hbar {\bf A}_n$.
In other words, the change of the momentum operator
\begin{eqnarray}
-i \hbar \nabla_{\bf R}  \rightarrow  -i\hbar \nabla_{\bf R}+\hbar {\bf A}_n  
\end{eqnarray}
occurs due to the presence of other particles.

Now, we shall consider a wave function for $N$ electron system,
\begin{eqnarray}
\Phi ({\bf r}_1, \cdots, {\bf r}_{N},t)
\end{eqnarray}
where ${\bf r}_j$ is the coordinate of the $j$th particle.

The Berry connection ${\bf A}_n$ takes into account the effect of the electron dynamics on the nuclear dynamics through the wave function they share
(i.e.,  $f({\bf R}) \langle {\bf r}|n({\bf R}) \rangle$).

Let us calculate the Berry connection arising from $\Phi ({\bf r}_1, \cdots, {\bf r}_{N},t)$ by regarding ${\bf r}_1$ as an adiabatic parameter.
In other words, we calculate the gauge field that affects the dynamics on the particle $1$ generated by other particles $2, \cdots, {N}$
through the wave function they share (i.e.,  $\Phi ({\bf r}_1, \cdots, {\bf r}_{N},t)$).

We cannot consider  ${\bf r}_1$ as a slow variable compared with other ${\bf r}_2, \cdots, {\bf r}_{N}$; however, the whole system stays in the same state described by the wave function $\Phi ({\bf r}_1, \cdots, {\bf r}_{N},t)$.

We first construct a normalized wave function $|n_{\Phi}({\bf r}_1) \rangle$ that contains ${\bf r}_1$ as an adiabatic parameter,
 \begin{eqnarray}
\langle  {\bf r}_{2}, \cdots, {\bf r}_{N} |n_{\Phi}({\bf r}_1,t) \rangle = { {\Phi({\bf r}_1, {\bf r}_{2}, \cdots, {\bf r}_{N},t)} \over {|C_{\Phi}({\bf r}_1 ,t)|^{{1 \over 2}}}}
\end{eqnarray}
Here, $|C_{\Phi}({\bf r}_1,t)|$ is the normalization constant given by 
\begin{eqnarray}
|C_{\Phi}({\bf r}_1,t)|=\int d^3 r_{2} \cdots d^3 r_{N}\Phi({\bf r}_1, {\bf r}_{2}, \cdots)\Phi^{\ast}({\bf r}_1,  {\bf r}_{2}, \cdots)
\end{eqnarray}

The wave function $\langle  {\bf r}_{2}, \cdots, {\bf r}_N|n_{\Phi}({\bf r}_1) \rangle$ is normalized to $1$ with respect to integrals over ${\bf r}_2, \cdots, {\bf r}_{N}$, and single-valued with respect to ${\bf r}_1$. 
Then, the {\em Berry connection for many-body wave functions} is calculated as
 \begin{eqnarray}
{\bf A}^{\rm MB}_{\Phi}({\bf r}_1,t)=-i \langle n_{\Phi}({\bf r}_1,t) |\nabla_{{\bf r}_1} |n_{\Phi}({\bf r}_1,t) \rangle
\label{AMphi}
\end{eqnarray}

As in the Born-Oppenheimer case,  ${\bf A}^{\rm MB}_{\Phi}({\bf r}_1,t)$ will modify the momentum operator as  
\begin{eqnarray}
-i \hbar \nabla_{{\bf r}_1}  \rightarrow  -i\hbar \nabla_{{\bf r}_1}+\hbar {\bf A}^{\rm MB}_{\Phi}({\bf r}_1)
\label{eqFBerry}
\end{eqnarray}
The same modification also occurs for ${\bf r}_i, i=2\cdots, {N}$.

Let us consider the effect ${\bf A}^{\rm MB}_{\Phi}({\bf r}_1,t)$ using the path integral formalism of quantum mechanics \cite{koizumi2021b}.
According to the path integral formalism, a wave function is a sum of contributions from all paths each contributes an exponential whose phase is the classical action divided by $\hbar$ for the path in question \cite{Feynmanpath}.
For the system of charged particles and electromagnetic field, the classical action $S$ are composed of the following three terms, 
\begin{eqnarray}
S=S_1+S_2+S_3
\end{eqnarray}
where
\begin{eqnarray}
S_1=\sum_i { m \over 2} \int dt \ \dot{\bf r}_i^2
\end{eqnarray}
is the action for the particles, 
\begin{eqnarray}
S_2=-\int d^3r dt \ \left[
\rho \phi^{\rm em}({\bf r},t) -{1 \over c} {\bf j} \cdot {\bf A}^{\rm em}({\bf r},t) \right]
=-q\sum_i \int  dt \ \left[ \phi^{\rm em}({\bf r}_i,t) -{1 \over c} \dot {\bf r}_i \cdot {\bf A}^{\rm em}({\bf r}_i,t) \right]
\nonumber
\\
\end{eqnarray}
is the action for the interaction between the field and particles, and
\begin{eqnarray}
S_3={1 \over {8 \pi}}\int d^3r dt \ \left[ ({\bf E}^{\rm em})^2- ({\bf B}^{\rm em})^2 \right]
={1 \over {8 \pi}}\int d^3r dt \ \left[ \left(-\nabla \phi^{\rm em} -{1 \over c} {{\partial {\bf A}^{\rm em}} \over {\partial t}} \right)^2- \left(\nabla \times {\bf A}^{\rm em}\right)^2 \right]
\nonumber
\\
\end{eqnarray}
is the action for the field. Here, $\phi^{\rm em}$ and ${\bf A}^{\rm em}$ are the scalar and vector potentials for the electromagnetic field, respectively; $\rho$ and ${\bf j}$ are the electric charge and current densities, respectively.

In the following we consider the case where the electric field ${\bf E}^{\rm em}$ is absent, and only the magnetic field ${\bf B}^{\rm em}$ is present.

The Berry connection ${\bf A}_{\Phi}^{\rm MB}$ modifies the momentum operator by providing
a $U(1)$ field. As a consequence, the effective vector potential for the $U(1)$ field in the system becomes
\begin{eqnarray}
{\bf A}^{\rm eff}={\bf A}^{\rm em}+{\bf A}^{\rm fic}, \quad {\bf A}^{\rm fic}= -{{\hbar c} \over q} {\bf A}_{\Phi}^{\rm MB}
\label{Aeff}
\end{eqnarray}
Here, we have used the ``fictitious'' vector potential ${\bf A}^{\rm fic}$ in place of ${\bf A}_{\Phi}^{\rm MB}$.

As will be shown later, ${\bf A}^{\rm fic}$ is given by
\begin{eqnarray}
{\bf A}^{\rm fic}=-{{\hbar c} \over {2e}} \nabla \chi
\label{Afic}
\end{eqnarray}
where $\chi$ is an angular variable with period $2 \pi$ \cite{koizumi2020}. 

By including the Berry connection, $S_2$ becomes,
\begin{eqnarray}
S_2'={1 \over c}\int d^3r dt \ {\bf j} \cdot {\bf A}^{\rm eff}({\bf r},t)
\label{eqS2'}
\end{eqnarray}
This gives rise to ``Lorentz force". The Lorenz force from ${\bf A}^{\rm fic}$ is zero in classical mechanics; however, 
${\bf A}^{\rm fic}$ may affect the dynamics of charged particles through the Aharonov-Bohm effect \cite{AB1959} in quantum mechanics.
Actually, this effect is the main concern of the present work. We call this term, the {\em Lorentz interaction} term.

The electromagnetic field energy is the energy stored in the space through the Lorentz interaction term.
If we use $S_2'$ as the Lorentz interaction term,
$S_3$ should be modified as
\begin{eqnarray}
S_3'=-{1 \over {8 \pi}}\int d^3r dt \ ({\bf B}^{\rm eff})^2 
\label{eqS3'}
\end{eqnarray}
 where ${\bf B}^{\rm eff}=\nabla \times{\bf A}^{\rm eff}$. 

${\bf B}^{\rm fic}=\nabla \times {\bf A}^{\rm fic}$ may not be zero due to the fact that $\chi$ may be multi-valued.

The use of $S_2'$ and $S_3'$, modify two of the Maxwell's equations,
\begin{eqnarray}
\nabla \cdot {\bf B}^{\rm eff}&=&0
\label{Maxwell1}
\\
\nabla \times {\bf B}^{\rm eff}&=&{{4 \pi } \over c}{\bf j}
\label{mMax2}
\end{eqnarray}

 The first one gives rise to a Dirac monopole as shown below: we rewrite it as
\begin{eqnarray}
\nabla \cdot {\bf B}^{\rm em}=-\nabla \cdot (\nabla \times {\bf A}^{\rm fic})
\label{eq71}
\end{eqnarray}
When the both sides of the above equation are integrated for a closed region with surface ${\rm Sf}$,
we have
\begin{eqnarray}
\int_{\rm Sf} d{\bf S} \cdot {\bf B}^{\rm em}=-\int_{\rm Sf} d{\bf S} \cdot (\nabla \times {\bf A}^{\rm fic})
\label{eq74}
\end{eqnarray}

We split ${\rm Sf}$ into two surfaces ${\rm Sf}_1$ and ${\rm Sf}_2$ with common boundary loop, $C=\partial ({\rm Sf}_1)=-\partial ({\rm Sf}_2)$. Then, we have 
the following relation
\begin{eqnarray}
\int_{{\rm Sf}_1} d{\bf S} \cdot (\nabla \times {\bf A}^{\rm fic})+\int_{{\rm Sf}_2} d{\bf S} \cdot (\nabla \times {\bf A}^{\rm fic})
=\int_{\partial( {\rm Sf}_1)} d{\bf r} \cdot {\bf A}^{\rm fic}+\int_{\partial( {\rm Sf}_1)} d{\bf r} \cdot {\bf A}^{\rm fic}
\nonumber
\\
\label{eq72}
\end{eqnarray}

We consider the case in which singularities exist in ${\bf A}^{\rm fic}$: let us take a closed surface $S$ with boundary $C=\partial S$, and ${\bf A}^{\rm fic}$ has a singularity in $S$.
Then, the following relation is obtained
\begin{eqnarray}
\int_{C} d{\bf r} \cdot {{\hbar c} \over {2e}} \nabla \chi ={{h c} \over {2e}} n 
\end{eqnarray}
where $n$ is an integer. 

If we have the case where $n=0$ for $\partial( {\rm Sf}_1)$ term and $n=1$ for $\partial( {\rm Sf}_2)$ term in Eq.~(\ref{eq72}), respectively,
Eq.~(\ref{eq74}) yields
\begin{eqnarray}
\int_{\rm Sf} d{\bf S} \cdot {\bf B}^{\rm em}={{h c} \over {2e}} 
\end{eqnarray}
This shows that a monopole with magnetic charge ${{h c} \over {2e}}$ exists in the region enclosed by ${\rm Sf}$ \cite{Monopole}. 

Let us show $\nabla \times {\bf B}^{\rm fic}=0$: 
it is well-known that $\nabla \chi$ in ${\bf A}^{\rm fic}$ can be decomposed as
\begin{eqnarray}
\nabla \chi=\nabla \chi_0 +\nabla f, \quad \nabla^2 \chi_0=0
\end{eqnarray}
where $f$ is a single-valued ($\nabla \times \nabla f=0$), and $\chi_0$ may be multi-valued. Thus, we have 
\begin{eqnarray}
\nabla \times {\bf B}^{\rm fic}=\nabla \times (\nabla \times \nabla \chi_0)= \nabla (\nabla^2 \chi_0)-\nabla^2 \nabla \chi_0=0
\label{Bfic0}
\end{eqnarray}

As a consequence Eq.~(\ref{mMax2}) is equal to 
\begin{eqnarray}
\nabla \times {\bf B}^{\rm em}&=&{{4 \pi } \over c}{\bf j}
\label{Maxwell2}
\end{eqnarray}

\section{Appearance of ${\bf A}^{\rm fic}$ from spin-twisting itinerant motion of electrons.}
\label{Appendix2}

The electrons in the normal state of the BCS superconductor are assumed to be 
Bloch electrons. We consider the Bloch electron under the influence of the Rashba interaction, and show that the spin-twisting itinerant motion is realized \cite{koizumi2020}.

 We use the following Rashba interaction term
\begin{eqnarray}
H_{so}= {\bm \lambda}({\bf r})\cdot {{\hbar{\bm \sigma}} \over 2} \times \left(\hat{\bf p}-{q}{\bf A}^{\rm em}({\bf r})\right), 
\end{eqnarray}
where ${\bm \lambda}({\bf r})$ is the spin-orbit coupling vector (its direction is the internal electric field direction), $\hat{\bf p}=-i\hbar \nabla$ is the momentum operator, and $q=-e$ is electron charge \cite{Rashba}. 

We take into account  the coordinate dependence of the spin function by adopting the following spin function 
\begin{eqnarray}
 \Sigma_1({\bf r})=e^{-{ i \over 2} \chi({\bf r})} 
 \left(
 \begin{array}{c}
 e^{-i { 1 \over 2} \xi ({\bf r})} \sin {{\zeta ({\bf r}) } \over 2}
 \\
  e^{i { 1 \over 2} \xi {\bf r})} \cos {{\zeta ({\bf r}) } \over 2} 
 \end{array}
 \right)
 \label{spin-d1}
\end{eqnarray}
for the Bloch electron,
where $\zeta ({\bf r})$ and $\xi ({\bf r})$ are the polar and azimuthal angles of the spin-direction, respectively.
The expectation value of spin ${\bf s}({\bf r})=(s_x({\bf r}), s_y({\bf r}), s_z({\bf r}))$ is given by
\begin{eqnarray}
s_x ({\bf r})= { \hbar \over 2} \cos \xi ({\bf r}) \sin \zeta ({\bf r}), \ s_y ({\bf r})= { \hbar \over 2} \sin \xi ({\bf r}) \sin \zeta ({\bf r}), \ s_z= { \hbar \over 2}  \cos \zeta ({\bf r})
\end{eqnarray}

The Berry connection arising from the spin function $\Sigma_1({\bf r})$ is
\begin{eqnarray}
{\bf A}^{\rm fic}_1({\bf r})=-i {{\hbar c}  \over e}\Sigma_1^{\dagger} \nabla \Sigma_1= -{{\hbar c} \over {2e}} \nabla \chi ({\bf r}) +{ {\hbar c} \over {2e}} \nabla \xi ({\bf r}) \cos \zeta ({\bf r})
\end{eqnarray}
and the effective vector potential is
\begin{eqnarray}
{\bf A}_1^{\rm eff} ={\bf A}^{\rm em}+ {\bf A}_1^{\rm fic}
\label{Aeff}
\end{eqnarray}

We introduce the following wave vector ${\bf k}_c$ for convenience, 
\begin{eqnarray}
{\bf k}_c = {\bf q}_c+{ e \over {\hbar c}} {\bf A}^{\rm eff}_1({\bf r}_c)
\label{gaugek}
\end{eqnarray}

By following the standard method for the wave packet formalism\cite{Niu,koizumi2020}, the following equations of motion are obtained,
\begin{eqnarray}
\dot{\bf r}_c&=&{ 1 \over \hbar} {{\partial {\cal E}} \over {\partial {\bf k}_c}}+ {\bm \lambda}({\bf r}_c) \times {\bf s}({\bf r}_c)
\label{eqm1}
\\
\dot{\bf k}_c&=& -{e \over {\hbar c }}\dot{\bf r}_c\times {\bf B}^{\rm eff},
\label{eqm2}
\end{eqnarray}
where ${\bf B}^{\rm eff}$ is the effective magnetic field,
\begin{eqnarray}
{\bf B}^{\rm eff}=\nabla \times {\bf A}_1^{\rm eff}
\end{eqnarray}

Eqs.~(\ref{eqm1}) and (\ref{eqm2}) indicate that the cyclotron motion in the band with energy
\begin{eqnarray}
{\cal E}({\bf k})+\hbar{\bm \lambda}({\bf r})\times {\bf s}({\bf r})\cdot {\bf k}
\label{NewB}
\end{eqnarray}
occurs.

By following the Onsager's argument\cite{Onsager1952}, let us quantize the cyclotron orbit. From Eq.~(\ref{gaugek}), the Bohr-Sommerfeld relation becomes
\begin{eqnarray}
\oint_C \hbar {\bf q}_c\cdot d{\bf r}_c=\oint_C (\hbar {\bf k}_c -{e \over  c} {\bf A}_1^{\rm eff}) \cdot d{\bf r}_c =2\pi \hbar \left(n+{ 1 \over 2} \right)
\label{Onsager1}
\end{eqnarray}
where $n$ is an integer and $C$ is the closed loop that corresponds to the section of Fermi surface enclosed by the cyclotron orbit.

Using Eq.~(\ref{eqm2}), we have 
\begin{eqnarray}
\oint_C \hbar {\bf k}_c \cdot d{\bf r}_c&=&-{e \over c} \oint_C d{\bf r}_c \cdot {\bf r}_c \times {\bf B}^{\rm eff}
\nonumber
\\
&=&{e \over c} \oint_C  {\bf B}^{\rm eff} \cdot  {\bf r}_c \times d{\bf r}_c
\nonumber
\\
&=&{{2e} \over c} \oint_C  {\bf A}_1^{\rm eff} \cdot d{\bf r}_c
\end{eqnarray}

Thus, Eq.~(\ref{Onsager1}) becomes
\begin{eqnarray}
{e \over c} \oint_C  {\bf A}^{\rm em} \cdot  d{\bf r}_c+{e \over c} \oint_C  {\bf A}_1^{\rm fic} \cdot  d{\bf r}_c=2\pi \hbar \left(n+{ 1 \over 2} \right)
\label{Onsager2}
\end{eqnarray}
This is the well-known quantized condition for the cyclotron motion if ${\bf A}_1^{\rm fic}$ is absent.

If ${\bf A}_1^{\rm fic}$ is present,
the above quantization condition is satisfied even if the magnetic field is absent. 
In this case, Eq.~(\ref{Onsager2}) yields,
\begin{eqnarray}
- \oint_C {1 \over 2} \nabla \chi ({\bf r})\cdot  d{\bf r}_c + \oint_C { 1 \over 2} \nabla \xi ({\bf r}) \cos \zeta ({\bf r}) \cdot  d{\bf r}_c=2\pi \left(n+{ 1 \over 2} \right)
\end{eqnarray}
which can be satisfied by the following two sets of conditions; one is $\zeta={\pi \over 2}$, $w_C[\chi]=-1$, and $n=0$; and the other is $\zeta={\pi \over 2}$, $w_C[\chi]=1$, and $n=-1$, where
\begin{eqnarray}
w_C[\chi]= {1 \over {2\pi}} \oint_C\nabla \chi ({\bf r})\cdot  d{\bf r}
\end{eqnarray}
is the winding number of $\chi$ along loop $C$. 

When $\zeta={\pi \over 2}$, the single-valued condition for $\Sigma_1({\bf r})$ in Eq.~(\ref{spin-d1}) leads to the single-valuedness of
$e^{-{ i \over 2} \chi({\bf r})} 
 e^{\pm i { 1 \over 2} }$. This is satisfied if 
the following requirement
\begin{eqnarray}
w_C[\chi]+w_C[\xi]= \mbox{even number}
\label{wcond}
\end{eqnarray}
is satisfied. 

The condition $w_C[\chi]=\pm1$ requires that $w_C[\xi]$ must be odd, thus, $w_C[\xi]$ is not zero. The nonzero value of $w_C[\xi]$ means that electrons perform spin-twisting itinerant motion. This indicates that the quantized cyclotron motion occurs without an external magnetic field by performing the spin-twisting itinerant motion.

When the quantized cyclotron motion without magnetic field occurs, ${\bf A}_1^{\rm fic}$ is given by
 \begin{eqnarray}
{\bf A}^{\rm fic}_1({\bf r})= -{{\hbar c} \over {2e}} \nabla \chi ({\bf r}) 
\end{eqnarray}

In addition to $\Sigma_1$ in Eq.~(\ref{spin-d1}), we consider the spin function $\Sigma_2$  that is orthogonal to $\Sigma_1$,
\begin{eqnarray}
 \Sigma_2({\bf r})=e^{-{ i \over 2} \chi({\bf r})} 
 \left(
 \begin{array}{c}
 ie^{-i { 1 \over 2} \xi ({\bf r})} \cos {{\zeta({\bf r})} \over 2}
 \\
 -i e^{i { 1 \over 2} \xi {\bf r})} \sin {{\zeta({\bf r})} \over 2}
 \end{array}
 \right)
 \end{eqnarray}

The fictitious vector potential from $\Sigma_2$ is 
\begin{eqnarray}
{\bf A}_2^{\rm fic}({\bf r})=-i {{\hbar c} \over e}\Sigma_2^{\dagger} \nabla \Sigma_2= -{{\hbar c} \over {2e}} \nabla \chi ({\bf r}) -{{ \hbar c} \over {2e}} \nabla \xi ({\bf r}) \cos \zeta ({\bf r})
\end{eqnarray}
 and the effective vector potential is
 \begin{eqnarray}
{\bf A}_2^{\rm eff} ={\bf A}^{\rm em}+ {\bf A}_2^{\rm fic}
\end{eqnarray}

When the quantized cyclotron motion without magnetic field occurs, ${\bf A}_2^{\rm fic}$ is given by
 \begin{eqnarray}
{\bf A}^{\rm fic}_2({\bf r})= -{{\hbar c} \over {2e}} \nabla \chi ({\bf r}) 
\end{eqnarray}
just like ${\bf A}_1^{\rm fic}$
 
 Let us consider the case where the many-electron wave function $\Phi$ for $N$ electrons is given as a Slater determinant
of spin-orbitals $\phi_{1,1}({\bf r})\Sigma_1({\bf r}), \phi_{1,2}({\bf r})\Sigma_2({\bf r})$, $\dots$, $\phi_{{N \over 2},1}({\bf r})\Sigma_1({\bf r})$, and $\phi_{{N \over 2},2}({\bf r})\Sigma_2({\bf r}) $,
where $\phi_{j,1}({\bf r})$ and $\phi_{j,2}({\bf r})$ are time-reversal partners and $N$ is assumed to be even. 

Then, ${\bf A}^{\rm MB}_{\Phi}$ in Eq.~(\ref{AMphi}) is calculated as
\begin{eqnarray}
 {\bf A}^{\rm MB}_{\Phi}&=& \  \rm{Im}
{ { \sum_{j=1}^{N \over 2} \left[ \phi^{\ast}_{j,1}({\bf r})\Sigma^{\dagger}_1({\bf r}) \nabla \phi_{j,1}({\bf r})\Sigma_1({\bf r})+\phi^{\ast}_{j,2}({\bf r})\Sigma^{\dagger}_2({\bf r})\nabla 
 \phi_{j,2}({\bf r})\Sigma_2({\bf r}) \right]} \over
  { \sum_{j=1}^{N \over 2} \left[ \phi^{\ast}_{j,1}({\bf r}) \phi_{j,1}({\bf r})+\phi^{\ast}_{j,2}({\bf r})
 \phi_{j,2}({\bf r}) \right] } }
 \nonumber
 \\
 &=&
 \  { e \over \hbar}
{ {{\bf A}^{\rm fic}_1\sum_{j=1}^{N \over 2}\phi^{\ast}_{j,1}({\bf r})
 \phi_{j,1}({\bf r}) + {\bf A}^{\rm fic}_2\sum_{j=1}^{N \over 2}\phi^{\ast}_{j,2}({\bf r})
 \phi_{j,2}({\bf r})} \over
  { \sum_{j=1}^{N \over 2} \left[ \phi^{\ast}_{j,1}({\bf r}) \phi_{j,1}({\bf r})+\phi^{\ast}_{j,2}({\bf r})
 \phi_{j,2}({\bf r}) \right] } }
\end{eqnarray}
where ``Im'' indicates the imaginary part. Note that 
\begin{eqnarray}
\sum_{j=1}^{N \over 2} \left[ \phi^{\ast}_{j,1}({\bf r}) \nabla \phi_{j,1}({\bf r})+\phi^{\ast}_{j,2}({\bf r})\nabla \phi_{j,2}({\bf r}) \right]  \mbox{ is real }
\end{eqnarray}
since $\phi_{j,1}({\bf r})$ and $\phi_{j,2}({\bf r})$ are time-reversal partners.

When the quantized motion of spin-twisting itinerant motion occurs, ${\bf A}^{\rm fic}_1$ and ${\bf A}^{\rm fic}_2$ are given by ${\bf A}^{\rm fic}_1={\bf A}^{\rm fic}_2=-{ {\hbar c}\over {2e}} \nabla \chi$.
 Thus, we have
\begin{eqnarray}
 {\bf A}^{\rm MB}_{\Phi}=-{ 1 \over 2} \nabla \chi
 \label{AMB}
\end{eqnarray}

It is shown that when the pairing energy gap is formed, the smallest possible size of the loop current from the quantized motion is estimated to be $\xi_{\rm BCS}$  in Eq.~(\ref{xi-BCS}) \cite{koizumi2021b}. The loop currents can coexist with displacing their centers.
This suggests that $\xi_{\rm BCS}$-sized loop currents exist in the BCS superconductor with the current directions arranged in such a way that 
the macroscopic current is zero.

\section{Reversible superconducting-normal metal phase transition in a magnetic field.}
 \label{Appendix3}

We will show that how the existence of ${\bf A}^{\rm fic}$ explains the reversible superconducting-normal metal phase transition in a magnetic field.

 Let us consider the magnetic field part of the free energy of the Ginzburg-Landau theory is given by
\begin{eqnarray}
 F_{\rm mag}=\int d^3 r {1 \over {8\pi}} \left( {\bf B}^{\rm eff}  \right)^2 
  \label{mag}
 \end{eqnarray}
 
 During the phase transition, the superconducting region shrinks and the magnetic field penetrating region increases.
 As a consequence $\partial_t {\bf B}^{\rm em} \neq 0$ occurs. This will induce an electric field, and if normal current exists, the Joule heating should occur.
 Thus, for the reversible transition to occur, it should occur without generating normal current.
 
 The change of  $F_{\rm mag}$ due to $\partial_t {\bf B}^{\rm eff} \neq 0$ is given by
 \begin{eqnarray}
 \Delta F_{\rm mag}=\int d^3 r \int_t^{t+\Delta t} dt{1 \over {4\pi}} {\bf B}^{\rm eff} \cdot \partial_t {\bf B}^{\rm eff} 
 \end{eqnarray}
 where $\Delta t$ is the time interval for this change
 
  If the above change of ${\bf B}^{\rm eff}$ is only due to the change of ${\bf A}^{\rm fic}$, $\Delta F_{\rm mag}$ is given by
   \begin{eqnarray}
 \Delta F_{\rm mag}={1 \over c}\int d^3 r {\bf j}_s \cdot \int_t^{t+\Delta t} dt \ \partial_t {\bf A}^{\rm fic} 
  \label{eq212}
 \end{eqnarray}
 where 
    \begin{eqnarray}
    {\bf B}^{\rm eff}=\nabla \times ({\bf A}^{\rm em}+{\bf A}^{\rm fic}), \quad {\bf j}_s={ c \over {4\pi}} \nabla \times {\bf B}^{\rm eff}
 \end{eqnarray}
 are used.
 
 The supercurrent kinetic energy part of the free energy is given by
 \begin{eqnarray}
 F_{\rm kin}=\int d^3 r {m_e \over {2}}n_s{\bf v}^2=\int d^3 r {{e^2 n_s} \over {2m_e c^2}} \left( {\bf A}^{\rm eff} \right)^2
\end{eqnarray}
where the supercurrent density 
    \begin{eqnarray}
    {\bf j}_{s}=-en_s {\bf v}=-{ {e^2 n_s} \over {m_e c}}{\bf A}^{\rm eff}
 \end{eqnarray}
 is used.

 The change of the supercurrent kinetic energy $\Delta F_{\rm kin}$ caused by $\partial_t {\bf A}^{\rm fic}$ is
 \begin{eqnarray}
 \Delta F_{\rm kin}=-{1 \over c}\int d^3 r {\bf j}_s \cdot \int_t^{t+\Delta t} dt \ \partial_t {\bf A}^{\rm fic} 
 \label{eq216}
\end{eqnarray}

Therefore, according to Eqs.~(\ref{eq212}) and  (\ref{eq216}), the transition with $ \Delta F_{\rm mag}+ \Delta F_{\rm kin}=0$ is realized via 
 \begin{eqnarray}
 \int_t^{t+\Delta t} dt \ \partial_t {\bf A}^{\rm fic} 
\end{eqnarray}
This occurs in a quantized manner by changing the winding numbers for $\chi$, without Joule heating.
This is the key step to realize the reversible superconducting-normal metal phase transition in a magnetic field.
Other changes occur among ${\bf j}_s$, ${\bf B}^{\rm em}$, $n_s$, 
to satisfy $\nabla {\bf B}^{\rm em}={ {4\pi} \over c}{\bf j}_s$ and $\partial_t n_s +\nabla \cdot {\bf j}_s=0$.
However, those changes can proceed without Joule heating \cite{koizumi2020b}. 

\section{Collective mode $\chi$ and associated number changing operators}
 \label{Appendix4}

We consider the quantization of the collective mode arising from $\chi$. For this purpose, we use
the Heisenberg operator formalism.

In order to have the phase arising from ${\bf A}^{\rm fic}$ in Eq.~(\ref{eqS2'}),
we express $\Phi$ as
\begin{eqnarray}
\Phi=\Phi_0 e^{ -{ i \over 2} \sum_{j=1}^N \chi({\bf r}_j)}
\end{eqnarray}
by explicitly extracting the Berry phase part $e^{ -{ i \over 2} \sum_{j=1}^N \chi({\bf r}_j)}$.

This is actually the definition for $\Phi_0$ through $\Phi$ and the Berry connection,
\begin{eqnarray}
\Phi_0=\Phi e^{ { i \over 2} \sum_{j=1}^N \chi({\bf r}_j)}
\end{eqnarray}

The kinetic energy part of the Hamiltonian is given by
\begin{eqnarray}
K_0={ 1\over {2m_e}} \sum_{j=1}^{N} \left( {\hbar \over i} \nabla_{j} \right)^2
\end{eqnarray}
where $m_e$ is the electron mass. The current density calculated using the above kinetic energy
\begin{eqnarray}
{\bf j}({\bf r})=-q \sum_{j=1}^N {\hbar \over {2 im_e}} \int d^3 r_1 \cdots d^3 r_N \
\left[
\Phi_0^{\ast} \nabla_j \Phi_0 - \Phi_0 \nabla_j \Phi_0^{\ast} \right] \delta({\bf r}-{\bf r}_j) 
\end{eqnarray}
 is zero; i.e., $\Phi_0$ is a currentless state.

When the ground state does not have the non-trivial Berry connection, it is a currentless state.
In this case, we may identify $\Phi$ as $\Phi_0$, and the collective coordinate $\chi$ disappears.

However, when the Berry connection is non-trivial, the collective coordinate $\chi$ is present.
We treat $\chi$ as a collective dynamical variable, and quantize it with the help of the following Lagrangian,
\begin{eqnarray}
{\cal L}\!=\langle \Phi | i\hbar \partial_t \!-\!H| \Phi \rangle\!=\! i\hbar \langle \Phi_0 | \partial_t | \Phi_0 \rangle+{\hbar \over 2} \int \!d{\bf r} \ {{n_p \dot{\chi} }} - \langle \Phi |H| \Phi \rangle
\label{L}
\end{eqnarray}
where $n_p$ is the number density of the particles \cite{Koonin1976}.

From the above Lagrangian, the conjugate momentum of $\chi$ is obtained as
\begin{eqnarray}
p_{\chi}= {{\delta {\cal L}} \over {\delta \dot{\chi}}}={\hbar \over 2} n_p
\label{momentumchi}
\end{eqnarray}
thus, $\chi$ and $n_p$ are canonical conjugate variables.

By following the canonical quantization condition 
\begin{eqnarray}
[\hat{p}_{\chi}({\bf r}, t), \hat{\chi}({\bf r}', t)]=-i\hbar \delta ({\bf r}- {\bf r}')
\end{eqnarray}
where $\hat{p}_{\chi}$ and $\hat{\chi}$ are operators corresponding to ${p}_{\chi}$ and ${\chi}$ respectively, 
we obtain the following relation
\begin{eqnarray}
\left[{ {\hat{n}({\bf r}, t)} } , \hat{\chi}({\bf r}', t) \right]=-2i \delta ({\bf r}- {\bf r}')
\label{commu0}
\end{eqnarray}
where $\hat{n}$ is the operator corresponding to $n_p$.

Strictly speaking, $\hat{\chi}$ is not a hermitian operator; however, it is known that when it is used as $e^{\pm {i \over 2}{\hat{\chi} }}$, the problem is avoided, practically \cite{Phase-Angle}. 
  
We construct the following boson field operators from $\hat{\chi}$ and $\hat{n}$,
\begin{eqnarray}
\hat{\psi}^{\dagger}({\bf r})= \left(\hat{n}({\bf r}) \right)^{1/2} e^{{i \over 2} {\hat{\chi}({\bf r}) }}, \quad \hat{\psi}({\bf r})= e^{-{i \over 2}{ \hat{\chi}({\bf r})}}\left( \hat{n}({\bf r}) \right)^{1/2}
\label{boson1}
\end{eqnarray}

Using Eq.~(\ref{commu0}), the following relations are obtained,
\begin{eqnarray}
 [\hat{\psi}({\bf r}),\hat{\psi}^{\dagger}({\bf r}')]=\delta({\bf r}-{\bf r}'), \quad  [\hat{\psi}({\bf r}),\hat{\psi}({\bf r}')]=0, \quad  [\hat{\psi}^{\dagger}({\bf r}),\hat{\psi}^{\dagger}({\bf r}')]=0
 \label{commu2}
\end{eqnarray}
  
From Eq.~ (\ref{commu2}), the following relations are obtained 
   \begin{eqnarray}
[e^{\pm {i \over 2} \hat{\chi}({\bf r})}, \hat{\psi}^{\dagger}({\bf r}')\hat{\psi}({\bf r}')]=\mp e^{ \pm {i \over 2} \hat{\chi}({\bf r})}\delta({\bf r}-{\bf r}')
\label{commchi}
 \end{eqnarray}
 which indicate that $e^{ \pm {i \over 2}  \hat{\chi}({\bf r})}$ are number changing operators for the number density operator
    \begin{eqnarray}
 \hat{n}({\bf r})=\hat{\psi}^{\dagger}({\bf r})\hat{\psi}({\bf r}) 
  \end{eqnarray}



%

\end{document}